A DETERMINATION OF THE FINE STRUCTURE CONSTANT USING PRECISION

MEASUREMENTS OF HELIUM FINE STRUCTURE

Marc Smiciklas, B.S., M.S.

Dissertation Prepared for the Degree of

DOCTOR OF PHILOSOPHY

UNIVERSITY OF NORTH TEXAS

August 2010

APPROVED:

David Shiner, Major Professor
Duncan Weathers, Committee Member
Dennis Mueller, Committee Member
Jose Perez, Committee Member
Chris Littler, Chair of the Department of
    Physics
James D. Meernik, Acting Dean of the
    Robert B. Toulouse School of
    Graduate Studies

Smiciklas, Marc. *A determination of the fine structure constant using precision measurements of helium fine structure*. Doctor of Philosophy (Physics), August 2010, 71 pp., 3 tables, 22 illustrations, 29 references.

Spectroscopic measurements of the helium atom are performed to high precision using an atomic beam apparatus and electro-optic laser techniques. These measurements, in addition to serving as a test of helium theory, also provide a new determination of the fine structure constant $\alpha$. An apparatus was designed and built to overcome limitations encountered in a previous experiment. Not only did this allow an improved level of precision but also enabled new consistency checks, including an extremely useful measurement in $^3$He. I discuss the details of the experimental setup along with the major changes and improvements. A new value for the J = 0 to 2 fine structure interval in the $2\,^3$P state of $^4$He is measured to be 31 908 131.25(30) kHz. The 300 Hz precision of this result represents an improvement over previous results by more than a factor of three. Combined with the latest theoretical calculations, this yields a new determination of $\alpha$ with better than 5 ppb uncertainty, $\alpha^{-1}$ = 137.035 999 55(64).

ACKNOWLEDGEMENTS

This research would not have been possible without the financial support of the National Science Foundation, the National Institute of Standards and Technology, and the University of North Texas. The interest that has been shown to this project by a wide range of people has been greatly appreciated.

I thank the faculty and the staff at the University of North Texas for all the ways that they contributed to this project. Most importantly, I thank Dr. David Shiner for his excellent guidance and support that has helped me throughout this work.

# TABLE OF CONTENTS

LIST OF TABLES

# LIST OF ILLUSTRATIONS

# CHAPTER 1

## INTRODUCTION

Precision measurements in physics play a crucial role in not only developing and testing the limits of advanced experimental techniques, but also testing current theory, and ultimately improving our understanding of the physical world. In this regard, the study of the helium atom is especially important. Helium theory is a testing ground for advanced quantum electro-dynamical (QED) calculations and variational techniques. This is because of the helium atom being the simplest multi-electron atom, and the fact that, unlike hydrogen, there is no exact analytical solution to the Schrodinger equation. One of the most compelling reasons to study the helium atom, specifically measuring the helium fine structure to very high precision, is as a determination of the fine structure constant $\alpha$. The fine structure constant is the fundamental constant of nature that governs the strength the electromagnetic interaction and is the only true adjustable parameter in QED theory. Though $\alpha$ has been very precisely determined through measurements of the electron magnetic moment [1], an independent method for an alternative determination is vital as a test of theory. To use atomic fine structure, the uncertainties must be at the level of 100 million times smaller or 10 parts per billion (ppb) than the measured interval in order to yield a competitive alternative value. Helium fine structure is ideally suited for this, owing to the recent advances in helium theory and being some 200 times more sensitive in determining $\alpha$ than hydrogen. For my dissertation, I have designed and built a complete experimental apparatus and used this setup to measure the 32 GHz J = 0 to J = 2 fine structure interval in the $2^3$P state of $^4$He with an uncertainty of 300 Hz. This is a 10 ppb uncertainty in the fine structure interval and yields a 5 ppb determination of $\alpha$ (since fine structure ~ $\alpha^2$).

The helium atom played an important role in the development of quantum mechanics, beginning with Werner Heisenberg's first variational calculations and the subsequent advances by Egil Hylleraas using explicit electron-electron separation coordinates [2]. These methods were subsequently employed and further developed by a number of major contributors. In 1964, the theorist Charles Schwartz suggested the use of helium fine structure theory and measurements as a precise determination of the fine structure constant [3]. This was in conjunction with the early experimental work done by Vernon Hughes [4]. Important advances in the calculation of helium fine structure were then carried out by Douglas and Kroll [5], Lewis and Serafino [6], Drake and coworkers [7], Sapirstein and Pachucki [8], among others. This has culminated in the most recently published calculations by Krzysztof Pachucki [9] where all terms in helium fine structure, up to order $m\alpha^7$, have been precisely evaluated.

In order to perform spectroscopic measurements of atomic energy levels, a photon of the necessary energy must be supplied to induce a transition. However, since fine structure levels in $2^3P$ state of helium are unstable, it is necessary to use some other "stable" state to reach those levels. For this purpose, the metastable $2^3S$ state is used. From here, the atoms can be excited into the $2^3P$ state and then excited again to undergo transition between the fine structure levels. This first method was used by the early Hughes experiments [4], as well as more modern applications of this technique by a group at York University [10,11]. A second approach, which is the technique used in this experiment, is to measure each fine structure energy level with respect to the metastable state and take the difference to determine the splittings. The fist technique is a direct microwave measurement, since the fine structure intervals in the $2^3P$ state correspond to 2.3 GHz up to 32 GHz in separation. The $2^3S$ state has a nearly ten thousand times larger separation at about 2.8 THz (1083 nm) from the $2^3P$ state (see Fig. 1). The second approach is a modern laser technique since it takes advantage of fairly recent advances in diode lasers at the 1083 nm wavelength. The laser approach has been used in previous incarnations of this experiment both here [12] and at Yale [13], as well as by another group at Harvard [14] and at LENS in Italy [15], although Harvard and LENS use very different implementations than the technique discussed here.

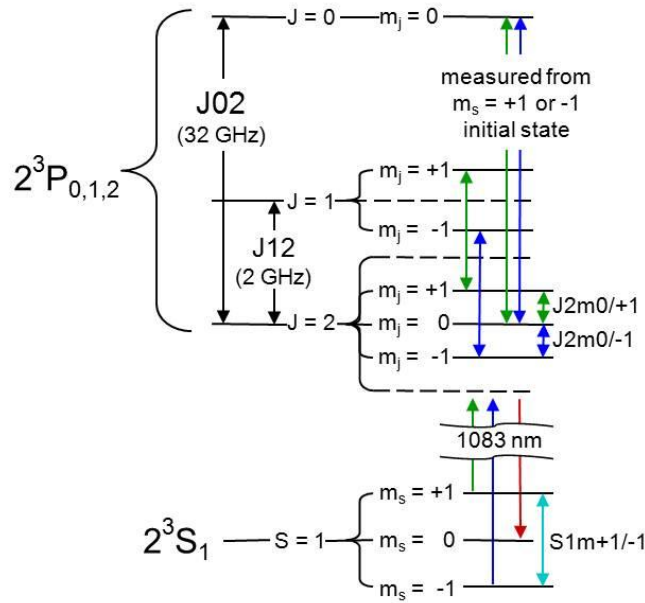

Fig. 1. $^{4}$He level diagram.

The experiment reported in this dissertation, uses an electro-optically tuned 1083 nm diode laser to excite helium atoms in an atomic beam. The apparatus has been redesigned and completely rebuilt from that used in a previous experiment [12]. I have incorporated major changes in design and function that significantly impact the performance and final results. After giving a brief theoretical introduction in Chapter 2, I discuss in Chapter 3 these changes along with each detail of the experimental setup. The analysis and final results for the experiment, along with a very important test using $^{3}$He, are presented in Chapter 4. Finally, Chapter 5 gives my concluding remakes.

CHAPTER 2

THEORETICAL FRAMEWORK

Introduction

In this chapter, I attempt to setup some of the standard formalism used to describe the helium atom. This is not only to establish a basic understanding of the helium atom and its structure, but also to present some very important theoretical procedures used to calculate useful, and even necessary, information. This includes being able to predict where to look for the individual transitions, what size signal is expect for those transitions, and most importantly, obtain required magnetic field corrections applied directly to the actual experimental results. Since $^3$He is used as an important check in this experiment, both the $^4$He and $^3$He isotopes are presented. Much of the information discussed here can be found in most introductory text books on quantum mechanics. For a more detailed discussion, I highly recommend the introductory text by Griffiths [16], or for a more advanced treatment, the graduate level text by Shankar [17].

The helium atom is a three body system consisting of two electrons and a nucleus. The nucleus contains two protons ($Z = 2$) and, for this experiment, either two neutrons for $^4$He or a single neutron for $^3$He. From here, it is useful to identify the quantum numbers of the system (i.e. degrees of freedom). These are the $n$, $l$, and $m$ values obtained by solving the Schrodinger equation. For the purposes of this experiment, the principle quantum number $n$ can be restricted to $n = 2$ since only transitions between the 2S and 2P states are considered. The azimuthal quantum number $l$ represents the orbital angular momentum of the electrons. The total angular momentum ($L$) of the system is the sum of the orbital angular momenta ($l$) of the electrons. For this experiment, one electron always remains in the ground state while the other is in the excited state. The possible values for the total orbital angular momentum are enumerated by

$$L = 0, \dots, n - 1.$$

Since $n = 2$, there are two possible total angular momentum states, $L = 0$ and $L = 1$. These correspond to the 2S and 2P states respectively. Before discussing the magnetic quantum number $m$, which is the spatial component of the system, additional degrees of freedom must be taken into account. Each electron has an intrinsic angular moment or spin, $s = \frac{1}{2}$. The total electron spin ($S$) of the system is given by

$$S = (s_1 - s_2), \dots, (s_1 + s_2).$$

Therefore, the total intrinsic spin has two possible values, $S = 0$ and $S = 1$. Before considering the total angular momentum of the system, let us first consider the 2S state where $L = 0$. Now, for the magnetic quantum numbers ($m_s$), we have the following standard relationship,

$$m_s = -S, \dots, +S.$$

This means that for $S = 0$, there is only a single state, $m_s = 0$; but for $S = 1$, there are three possible states, $m_s = $ -1, 0, and +1. These are the so called singlet ($S = 0$) and triplet ($S = 1$) states, named for obvious reasons. For the most part, we can focus on the triplet states since the singlet states are generally well separated in energy from the triplet states. This separation in energy comes from the antisymmetric spin configuration of the singlet state electrons feeling a larger repulsion and therefore energy shift than the more closely spaced symmetric spin configuration of the triplet states. Now, we can look at the spin-orbit interaction which is actually responsible for the fine structure splittings in the 2P state. So, for this state ($L = 1$), the total angular momentum ($J$) is given by

$$J = (S - L), \dots, (S + L).$$

Again, focusing on just the triplet states ($S = 1$), there are three possible total angular momentum values in the $2^3$P state, $J = 0$, 1, and 2. Like before, the magnetic quantum numbers ($m_j$) for each of the $J$ values are given by

$$m_j = -J, \dots, +J.$$

For a list of these and the other states discussed above as well as their relative positions, refer back to the diagram in Fig. 1. This completes the picture for $^4$He. However, $^3$He has an additional degree of freedom due to the nuclear spin not being equal to zero, $I = \frac{1}{2}$. This is what is responsible for the hyperfine structure in $^3$He. Following the same standard procedure for adding angular momentum quantum numbers as above, the grand total angular momentum ($F$) yields the following values…

For the $2^3$S state,

$$F = \frac{1}{2} \text{ and } \frac{3}{2}.$$

For the $2^3$P state,

$$J = 0 \Rightarrow F = \frac{1}{2},$$

$$J = 1 \Rightarrow F = \frac{1}{2} \text{ and } \frac{3}{2},$$

$$J = 2 \Rightarrow F = \frac{3}{2} \text{ and } \frac{5}{2}.$$

After enumerating all the $m_f$, there are exactly twice as many $^3$He levels are there are $^4$He, though not all of them are necessarily available. The primary selection rules limiting which transitions are available for this experiment are $\Delta m = 0$ or $\pm 1$ and $\Delta F = 0$ or $\pm 1$. This simply means that neither the angular momentum nor the magnetic moment can change by more than one unit. As a side note, the 2S states are metastable because of the selection rule $\Delta l = \pm 1$, forbidding the direct decay ($\Delta l = 0$). The singlet and triplet states do eventually decay by various processes, but the triplet states have a much longer lifetime because of the required spin flip due to the Pauli Exclusion Principle.

## Solutions to the Hamiltonian

Calculating detailed information about the helium atom is of importance when performing the experimental measurements. Not only are the locations of the transitions reliably predicted (which is especially important when the levels shift at different magnetic fields), but very useful information about the transition rates and probabilities can be extracted as well. This information is very valuable in understanding details such as how large the expected signal size should be, which states certain transitions are likely to decay into, which states recycle more than others, etc. However, the most important piece of information calculated is, very precisely, how the energies of the levels shift in a magnetic field. Those calculations are discussed in more detail later in the section on Zeeman level corrections, which are used to correct the final measured result to zero B-field.

Performing the calculations is usually a simple matter of determining the correct form of the Hamiltonian and evaluating it to determine the wavefunctions and their corresponding energies. For the helium atom, the Hamiltonian is

$$H = \left(-\frac{\hbar^2}{2m}\nabla_1^2 - \frac{1}{4\pi\epsilon_0}\frac{2e^2}{r_1}\right) + \left(-\frac{\hbar^2}{2m}\nabla_2^2 - \frac{1}{4\pi\epsilon_0}\frac{2e^2}{r_2}\right) + \frac{1}{4\pi\epsilon_0}\frac{e^2}{|\mathbf{r_1}-\mathbf{r_2}|} \ [16].$$

This is simply two hydrogen-like Hamiltonian operators with charge $2e$ (one for each electron due to the coulomb potential of the nucleus) and an additional term that describes the repulsion the two electrons feel from each other. Unfortunately, unlike the case of the hydrogen atom, there is no exact analytical solution to the Schrodinger equation with this Hamiltonian. Because of that additional electron-electron interaction, it becomes necessary to use one of a variety and quite often very advanced approximation techniques, depending on the desired accuracy.

For the purposes of the experiment, a phenomenological form of the Hamiltonian is used. In this form, the Hamiltonian is modeled on the interactions that are responsible for the energy shifts observed in the atom. The coefficients of the interaction terms determine the magnitude of their effect. These can either be previously calculated or empirically determined through observation of the actual transition energy levels. The phenomenological Hamiltonians for the 2S and 2P states are as follows,

$$\mathbf{H}_{2S} = \Delta S(\mathbf{s}_1 \cdot \mathbf{s}_2)$$

and

$$\mathbf{H}_{2P} = \Delta P(\mathbf{s}_1 \cdot \mathbf{s}_2) + E[(\mathbf{s}_1 + \mathbf{s}_2) \cdot \mathbf{L}],$$

where $\Delta S$ and $\Delta P$ are the coupling constants responsible for the spin-spin interaction and the observed singlet-triplet splitting, and $E$ is the coupling constant of the spin-orbit interaction that produces the actual fine structure splittings. By using the standard matrix formalism to express the spin and orbital angular momentum operators, the eigenvalues and eigenvectors for the Hamiltonian matrices can be found by solving the usual characteristic equation given by,

$$\mathbf{H}|\boldsymbol{\psi}\rangle = E|\boldsymbol{\psi}\rangle \quad \Rightarrow \quad (\mathbf{H} - E\boldsymbol{I})|\boldsymbol{\psi}\rangle = |\mathbf{0}\rangle,$$

where there is eigenvector ($\boldsymbol{\psi}$) for each level and a corresponding eigenvalue ($E$) which is the energy of that level. The dimensions of the Hamiltonian operator are equal to the number of levels. So, for the 2S state, there are three triplet states and one singlet state, which means $\mathbf{H}_{2S}$ is a 4 x 4 matrix which yields four eigenvectors and four eigenvalues. The 2P state's Hamiltonian ($\mathbf{H}_{2P}$) is a 12 x 12 matrix with nine triplet and three singlet states.

After the solutions to the Hamiltonian have been found and the eigenvectors for individual states identified, it is a relatively simple procedure to calculate the transition rates from the 2S to 2P states. This involves calculating the matrix elements of the laser perturbation $\langle \psi_f | \mathbf{p} \cdot \mathbf{A} | \psi_i \rangle$, or in the limit of the electric dipole approximation $\langle \psi_f | \mathbf{E} \cdot \mathbf{r} | \psi_i \rangle$, where $\psi_f$ and $\psi_i$ are the final and initial states respectively. Writing $\mathbf{E} \cdot \mathbf{r}$ in spherical coordinates (and disregarding the radial component and overall factors since they are the same for all transitions), we have the transition matrix

$$T_q = \int_0^{2\pi} \int_0^{\pi} Y_1^m(\theta, \phi) \sqrt{4\pi} \, Y_1^{-q}(\theta, \phi) Y_0^0(\theta, \phi) \sin \theta \, d\theta d\phi.$$

The complete Mathematica analysis to evaluate all the transition rates and energies can be found in Appendix A.

## Zeeman Level Corrections

In order to remove the degeneracy in each of the magnetic sublevels (Zeeman levels) and separate the transitions to avoid overlapping, the experiment is conducted in a uniform magnetic field. This naturally shifts the interval spacing as well. Therefore, it is necessary to apply very precise corrections to adjust the fine structure intervals back to the zero magnetic field value. Fortunately, the theory for the Zeeman level shifts is very well understood [18]. Any problem with the application of the theory is tested by measuring the fine structure intervals at a wide range of magnetic fields (up to 8 mT). This is discussed in detail Chapter 4 under the section Systematic Checks.

The magnetic field is measured during the experiment by collecting data on one or more of the splittings between the magnetic sublevels (refer back to in Fig. 1 Chapter 1 to see all the available Zeeman intervals). When the data are analyzed, the splitting is compared against the calculated Zeeman level shifts to determine a very accurate value for the B-field. This B-field value is then used to evaluate the other measured intervals. For this to be effective, it is necessary to calculate in advance the B-field shifts for all the transitions. Otherwise, matching the correct B-field to the observed frequency difference would be especially challenging and very time consuming. Using the procedure described in the previous section, the energy levels have been calculated for a range of B-field values. The calculations for each of the transitions have been fit to polynomials in orders of **B,** which are used for the actual magnetic field determination and corrections during the analysis. The quality of the fits for J = 0 to J = 2 and J = 1 to J = 2 fine structure intervals as well as the interval used to measure the $^3$He hyperfine splitting are shown in Fig. 2 (with actual polynomials used; the residuals are plotted).

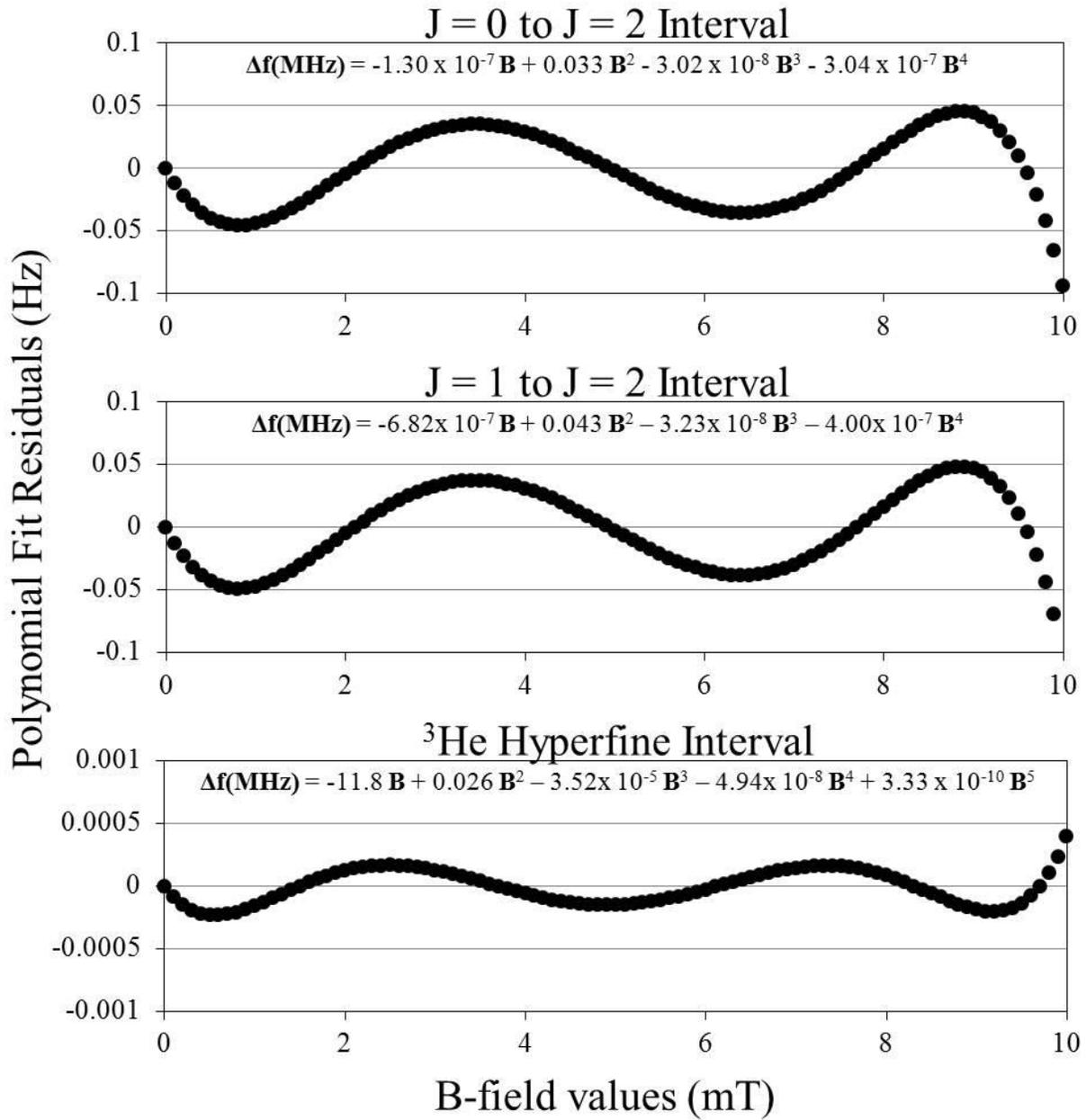

Fig. 2.  Plotted residuals to polynomial B-field fits (Δf).

CHAPTER 3

EXPERIMENTAL SETUP

Overview

Without a doubt, the biggest challenge of my dissertation and by far the most time consuming was redesigning the experimental setup almost entirely from the ground up. This new experimental setup incorporates new technologies and improved techniques over the old experimental setup that preceded it. The basic technique of using an electro-optic modulator (EOM) tuned 1083 nm diode laser to drive transitions in a metastable atomic beam is the same. However, key changes to the approach combined with the desire to build a cleaner apparatus necessitated a complete overhaul. This new experimental setup is superior in almost every way to its predecessor. The signal size has been greatly increased by a much improved metastable source design along with the incorporation of optical pumping. Also, the entire apparatus has been rebuilt using high grade stainless steel (mostly 316LN) for its very low magnetic properties and ultra-high vacuum (UHV) compatibility.

The apparatus in this experimental setup is a UHV chamber in which a highly collimated atomic beam (with Doppler width of 500 kHz) is initially prepared, undergoes laser interactions, and then is detected (see Fig. 3). The initial preparation starts with a metastable source to excite the helium atoms into the $m_s = +1, 0$, and $-1$ states. Immediately following the source, the atomic beam is prepared in the initial states by optically pumping the $m_s = 0$ states into the $m_s = \pm 1$ states. The prepared states are then driven in a uniform magnetic field by depolarizing laser excitations that repopulate the $m_s = 0$ detection state. Optionally, a mirror may be used to retro-reflect the interaction laser to cancel alignment sensitive Doppler effects. Before detection, the additional singlet state background created by the source is quenched using an electric field.

Finally, the $m_s = \pm 1$ states are deflected out of the beam, and the $m_s = 0$ states are detected using a channel electron multiplier. Each step in this process is discussed in the sections that follow.

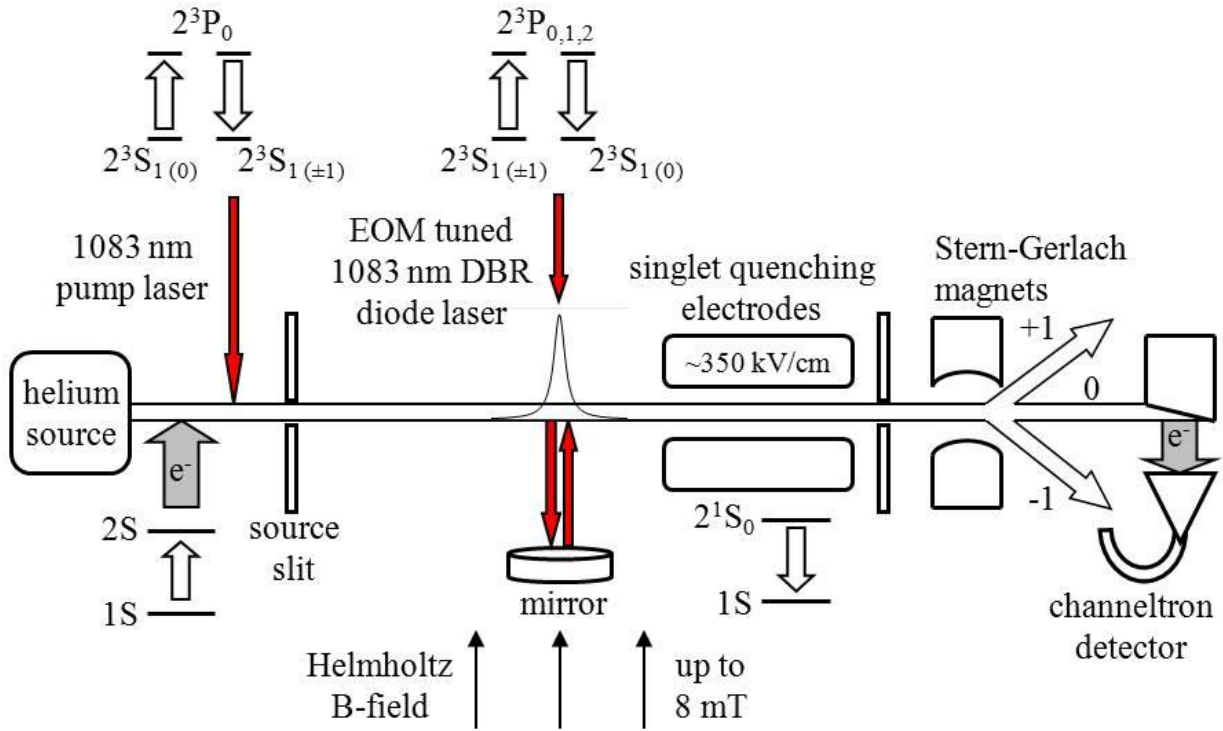

Fig. 3. Schematic diagram of the experimental apparatus.

Metastable Source

A great deal of time and effort was spent designing and testing the metastable helium source for this experimental setup. The efficiency of the source directly affects the size of the signal, and therefore, the final uncertainties in the measurements. When performing precision measurements, statistical uncertainties due to $\sqrt{N}$ counting noise are a major limiting factor. Reducing these uncertainties allows for shorter data runs and the ability to perform more systematic checks at the desired level of precision. My final design for the metastable helium source is substantially improved over that used in our previous experimental setup. This includes easier maintenance, increased reliability, and better signal strength. I discuss the basic operation as well as these improvements in greater detail below.

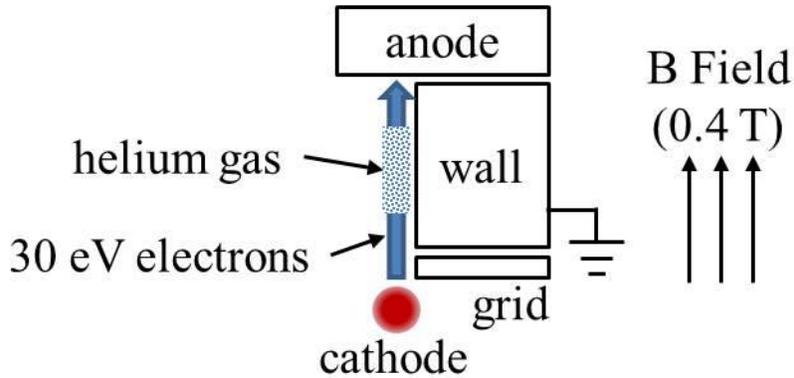

Fig. 4. Metastable source electron gun.

For this experiment, metastable helium in the atomic beam is generated by means of electron bombardment [19]. Electrons are pulled off a thermionic emitter [20] and cross at right angles to helium atoms in the atomic beam. The source arrangement is basically an electron gun which consists of a cathode to supply the electrons, a grid to accelerate the electrons to the necessary kinetic energy, a wall to reduce space charge effects, and an anode to collect the electrons (see Fig. 4). The cathode generates the electrons by effectively boiling them off a hot tungsten filament. A negative bias voltage between the cathode and the grid accelerate the electrons to approximately 30 eV of energy. The threshold for generating n = 2 metastable helium from the ground state is just below 20 eV. A large axial magnetic field supplied by two neodymium magnets sandwiching the source (~0.4 T) guides the electrons across the atomic beam and on to the anode. A grounded wall is placed very near the path the electrons travel. This is to provide a uniform potential for the electrons so as to maintain a constant kinetic energy when interacting with the helium atoms. Also, the positive image charges on the grounded wall serve to reduce space charge effects that limit the maximum electron emission from the filament.

The anode can be allowed to float by electrically disconnecting it, reflecting the electrons back across the helium atoms. Operating in reflection mode allows for significant gains in signal strength, as is discuss in more detail below.

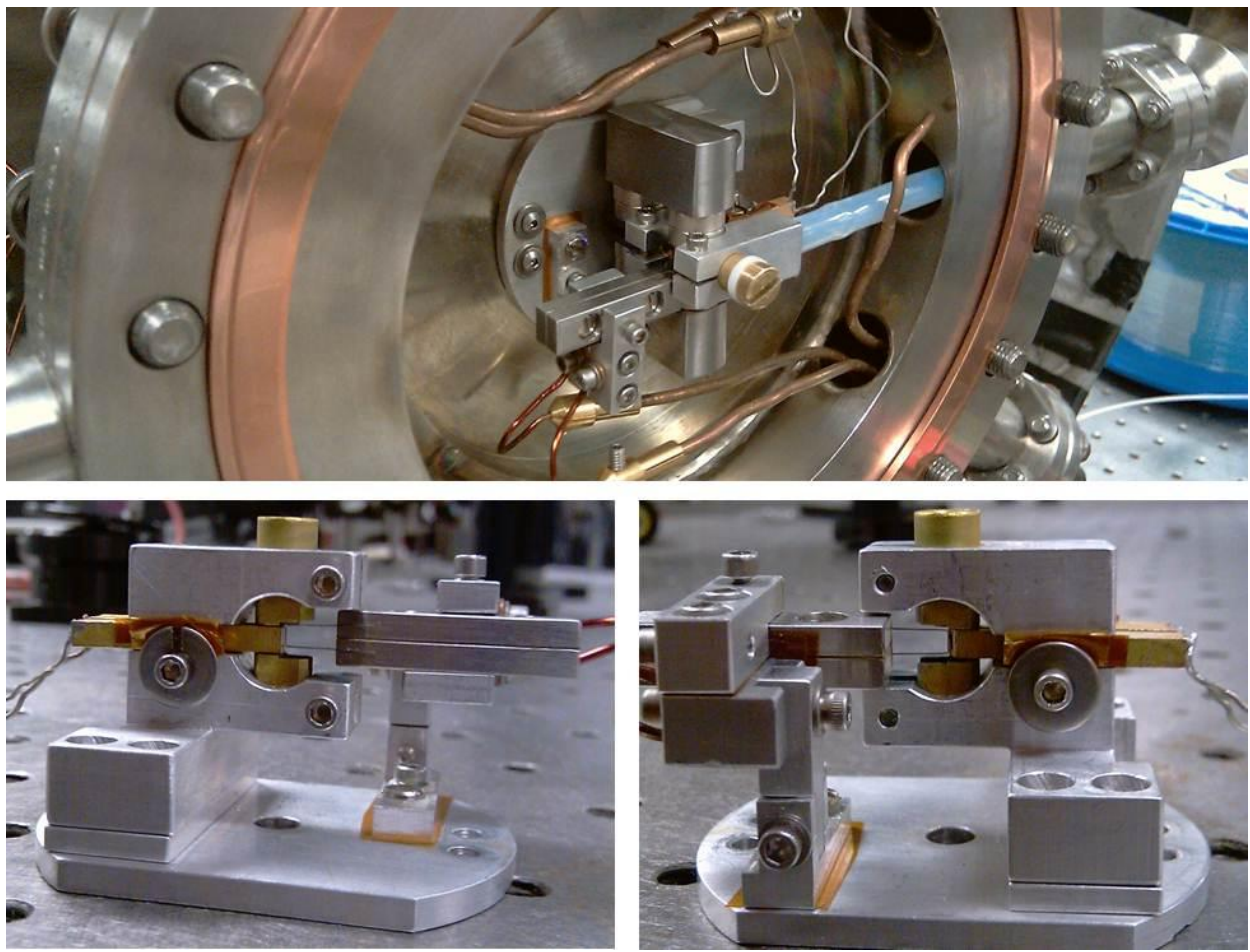

Fig. 5. Pictures of metastable helium source.

Before starting this project, I had gained considerable experience working with the previous experimental setup. This included regularly performing the tedious maintenance required when a filament needed to be replaced. With that experience in mind, my new source design (shown in Fig. 5) is significantly simpler to maintain. The source is very compact, totally contained on a single mount, and easily removed from the apparatus. Replacing the filament is as simple as loosening one screw to remove its mount. The filament is a 150 μm diameter tungsten wire bent to have a 2 mm tip. The ends are spot welded to two insulated stainless steel blocks with the tip protruding about 6.3 mm from the end of the mount. Due to residual stresses in the wire after spot welding, the filament is annealed in a test vacuum chamber using a 3-4 ampere heating current. Unfortunately I have found that upon annealing, the filament will twist well beyond the alignment tolerances of the source mount. As a clever solution to this problem, I slip a simple sleeve made from two small pieces of glass slide over the filament. The glass sleeve forces the filament to stay straight during annealing while still allowing it to get red hot.

The atomic beam is created by effusing helium gas through a 0.15 x 1.5 mm source slit which is then excited by the electron emission. The heating current for the source is typically around 3 amperes. The operating characteristics of the metastable source are shown in Fig. 6. Space charge effects are much more pronounced when operating with the anode disconnected. However, the increased signal from the reflected electrons is significant (almost eight times higher at low emission but still twice as high at high emission). Comparing this to the old experimental setup, typical operating emission was 10 mA with a signal of at best 150 thousand counts/sec (anode disconnected), depending on how well the filament was aligned. Operating continuously at 10 mA of emission, the lifetime of the filament would only be 2-3 weeks. For this new experimental setup, the source is usually operated at 3 mA of emission. Since filament life time becomes exponentially better at lower emissions, the average lifetime is now on the order of several months. Even at this lower emission, the signal is double that of the old experiment (300 thousand counts/sec). After the metastable helium is created, a skimmer separates the excess gas from the atomic beam. The atomic beam then passes from the source chamber to the lower pressure interaction chamber through the first collimating slit. The helium gas in the atomic beam as well as the excess gas in the source chamber is pumped back into the foreline and through a recirculation line. Contaminating gasses are cryogenically pumped by a liquid nitrogen cooled molecular sieve, and the helium gas is returned to the source to be reused. The typical operating pressure for each phase of the experiment is shown in Table 1.

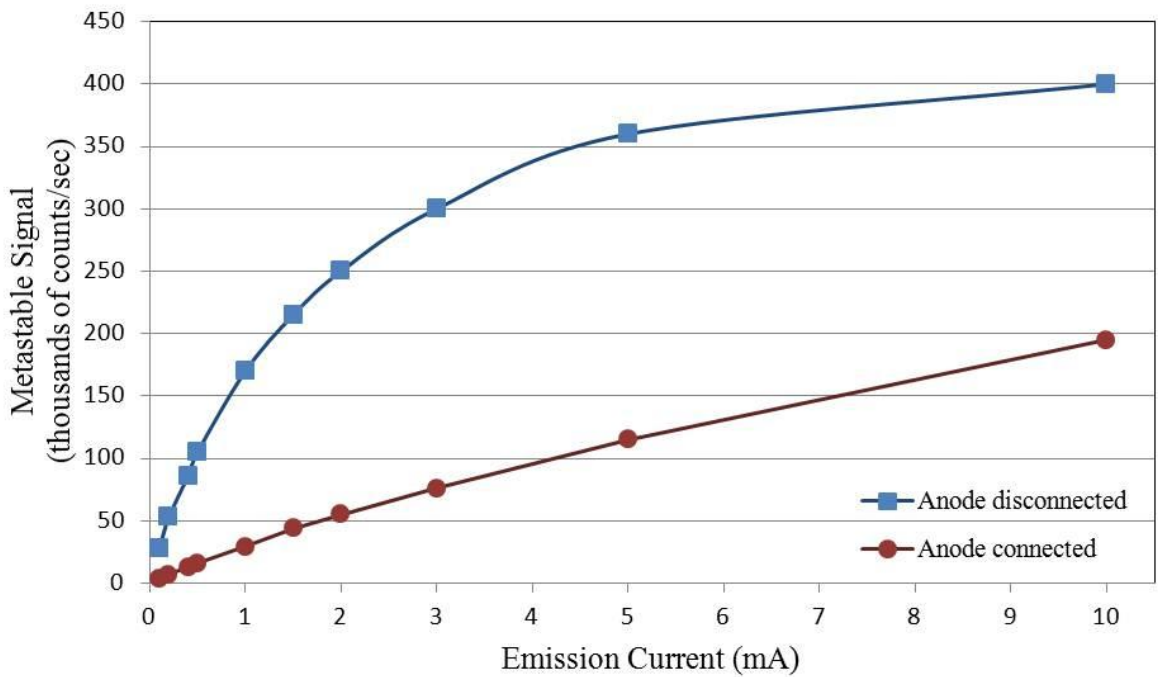

Fig. 6. Metastable source operating characteristics.

Table 1. Typical operating pressures.

| Location | Pressure (Torr) |
| --- | --- |
| Recirculation Line | 1 to 2 |
| Source Chamber | $(1 \text{ to } 10) \times 10^{-6}$ |
| Interaction Chamber | $(2 \text{ to } 4) \times 10^{-8}$ |

Optical Pumping

Before the atoms can be used in the laser interaction, the atomic beam must first be prepared into the initial state. Since $m_s = 0$ states are detected in order to determine if the laser is driving the transition, those states must first be removed from the beam so as not to contribute to the background. For that purpose, another laser is employed to optically pump the $m_s = 0$ state. Optically pumping also increase the signal in the $m_s = \pm 1$ states by 50% since the pumped atoms transfer into those states. A significant amount of research and testing went into building a custom designed Yb-doped fiber laser for this purpose.

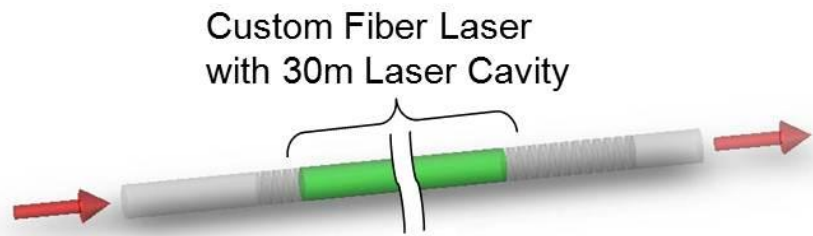

Fig. 7. Fiber laser diagram.

A fiber laser consists of a section of doped fiber (gain medium) spliced between two fiber Bragg gratings (front and back reflectors, see Fig. 7). A fiber coupled diode laser of the appropriate wavelength is used to pump the gain medium and start the lasing. The fiber used for the cavity gain is Liekki Yb1200 Ytterbium doped PM fiber. Ytterbium's peak absorption is centered at 976 nm (1200 dB/m for Yb1200 fiber), and it has a florescence tail that extends out to 1100 nm. The cavity length for the fiber laser is 30 m. A long cavity length is used to achieve maximum stability and ease of operation. This is because the fiber laser lases at multiple frequencies spaced by the free spectral range and has a bandwidth determined by the output grating (about 1 GHz for this laser). The free spectral range for a 30 m cavity length is approximately 3 MHz. Since the natural linewidth for 2S to 2P transition in helium is 1.6 MHz, the output of the fiber laser appears to be continuous. The center wavelength is controlled by temperature tuning the grating with a thermoelectric cooler. The 1 GHz bandwidth laser output is easily maintained on the pumping transition by stabilizing that temperature. The transition up to the $2^3P$ J = 0 level is used to pump the $m_s$ = 0 states (with 99.9% efficiency). Nearby transitions from the $m_s$ = ±1 states are moved well outside of the pump laser frequency range by pumping in a large magnetic field (~ 0.4 T). This is done by using two neodymium magnets located directly after the metastable source. Refer back to Fig. 5 to see the magnets for the optical pumping and the metastable source connected to each other with magnetic flux returns. The performance characteristics of the fiber laser are shown in Fig. 8.

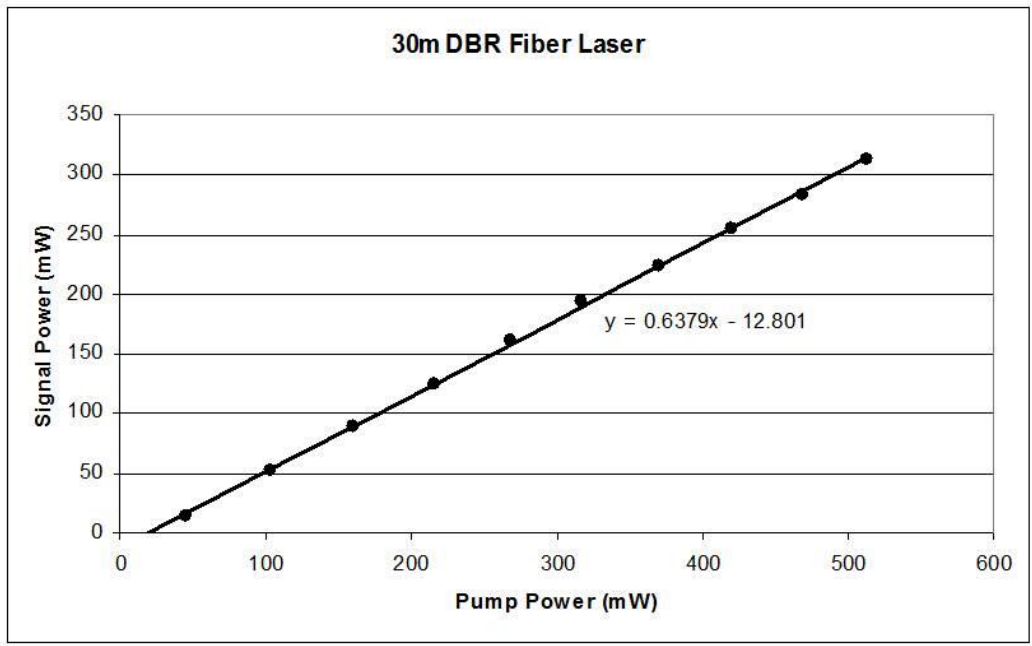

Fig. 8. Yb-doped fiber laser performance characteristics.

## Interaction Laser

The interaction laser used in this experimental setup is a 1083 nm distributed Bragg reflector (DBR) diode laser. The optical layout used for control and stabilization of the frequency and power of the laser is shown in Fig. 9. The output of the diode laser is couple into a fiber optic cable. Power stabilization is accomplished by tilting a piezo actuated mirror to vary the coupling efficiency into the fiber. A small amount of power is picked off and measured before the laser enters the apparatus. During the experiment, the intensity of the laser is controlled by changing the set point for the feedback electronics that control the piezos. Using this setup, the tuning range for the mirror gives about a factor of 8 in laser intensity control. The output frequency of the diode laser is locked to a reference frequency supplied by an iodine-stabilized HeNe laser. This locked and stabilized frequency serves as a carrier for an electro-optic modulator which creates tunable sidebands that are used to drive the transitions.

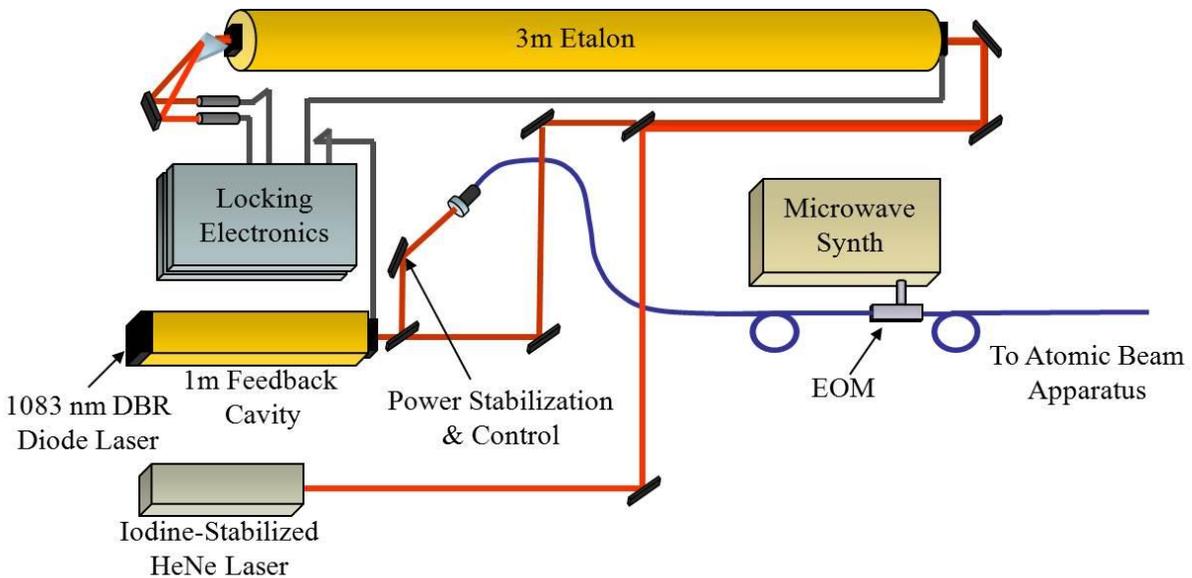

Fig. 9. Interaction laser optical layout.

During the frequency stabilization, the stability of an iodine-stabilized HeNe laser (made by Winter's Electro-Optics Inc.) is transferred to the 1083 nm diode laser. This is accomplished by using a resonant transfer cavity (3m Etalon in Fig. 9). The cavity is a Fabry-Perot interferometer (53 MHz free spectral range) with a piezo-actuated mirror on one end to control the length. A lock-in amplifier monitors the transmission through the interferometer and locks its length to an integer number of half wavelengths of the HeNe Laser. The diode laser is combined with the HeNe laser and passed through the interferometer. A second lock-in amplifier monitors the transmission and controls the length of a 1 m cavity feeding back into the diode laser. The feedback is tuned to a strong feedback regime to allow smooth control of the laser frequency and avoid frequency hops. The frequency of the diode laser is locked to a matching order number of the HeNe laser through the interferometer, i.e., the total length of integer number half wavelengths of the diode laser equals that of the HeNe laser.

Controlling the laser frequency during the experiment is accomplished by phase modulating the stabilized diode laser frequency. A GPIB (General Purpose Interface Bus) controlled microwave synthesizer couples the modulation frequency into an EOM which consists of a lithium niobate crystal. The crystal's index of refraction is modulated at the microwave frequency which provides the phase modulation for the diode laser. Phase modulation is very similar to frequency modulation since time dependence of the phase is simply added to the frequency, i.e.

$$\sin\left(\omega_c t + \varphi(t)\right)$$

where

$$\varphi(t) = \beta \sin\left(\omega_m t\right).$$

Here, $\omega_c$ is the carrier frequency, $\omega_m$ is the modulation frequency, and $\beta$ is the modulation index. For illustrative purposes, if we assume $\beta \ll 1$, we get the approximation

$$\sin(\omega_c t) + \beta \sin[(\omega_c \pm \omega_m)t].$$

Now we have three frequencies, the carrier frequency and frequencies above and below the carrier with the spacing exactly equal to the modulation frequency. If we allow any value for the modulation index (no longer limited to $\beta \ll 1$), this expression is represented as an infinite series yielding higher order sidebands with the amplitudes describe by Bessel functions. The EOM in this experimental setup has a 20 GHz bandwidth. In order to measure the 32 GHz fine structure splitting, the carrier is typically tuned roughly half way between the J = 0 and J = 2 levels in the $2^3$P state. The upper and lower sidebands are used to drive the transitions. To determine the fine structure interval, the difference between the modulation frequencies is taken.

Singlet State Quenching

When creating metastable helium in the source, both $2^3$S (triplet) and $2^1$S (singlet) states are produced. The singlet states have zero magnetic dipole moments ($m_s = 0$), and therefore do not deflect in a magnetic field. This means that singlet states are detected along with the triplet $m_s = 0$ states used to detect transitions. Also, singlet states are created in roughly equal numbers as the individual triplet states. This accounts for a very large contribution to the background (typically around 300 thousand counts/sec). In order to eliminate this very large background, the singlet states are quenched in an electric field.

The normal decay process for metastable singlet state atoms is spontaneous two-photon emission with a radiative lifetime of just under 20 ms [21]. Applying a large electric field mixes the $2^1$S state with higher $^1$P states. The $^1$P states very rapidly decay into the ground state [19], releasing a single photon.  Quenching the singlet states in an electric field is essentially stimulated two-photon emission in the limit of zero frequency for one of the photons [22]. Fortunately, the triplet states do not decay in this way because they require a spin flip from one of the electrons, which this cannot provide.  The rate at which the singlet states are quenched depends on the strength of the electric field and the time the atoms experience it.  The relationship can be expressed in the following form:

$$C = C_0 e^{-\gamma t}$$

where

$$\gamma = \gamma_0 + kE^2 \text{ [22]}.$$

Here, $\gamma$ is the total decay rate expressed in terms of the spontaneous decay rate $\gamma_0$, the quenching constant $k(= 0.933 \frac{cm^2}{kV^2.s})$ [22], and the electric field $E$ in kV/cm.  For the purpose of examining large quenching rates, we can ignore the spontaneous decay rate $\gamma_0$ due to its negligible contribution.

Considerable time and effort was spent designing the electrodes used to generate the quenching electric field.  Due to space requirements, the electrodes needed to be no longer than 10 cm.  With that limitation in mind, the electric field needed to be large enough to quench the singlet states at an acceptable rate.  The average velocity of the atomic beam is approximately 1600 m/s giving a maximum quenching time of about 60 μs.  Using the relationship above with this quenching time, the electric field should be about 350 kV/cm to achieve 99.9% quenching efficiency.  An electric field this large is susceptible to arcing and requires careful treatment.  To avoid breakdowns, the electrodes are made of hard anodized aluminum with an oxide layer approximately 50 μm thick [23].  However, another effect was observed that caused sporadic arcing even with the thick oxide layer.  Originally, the electrodes were mounted before the laser interaction close to the metastable source. Testing showed that arcing occurred when the metastable atoms were generated in the atomic beam. However, if the electron emission in the source is turned off, no arcing is observed. I was able to determine that metastable atoms colliding with the electrode walls precipitated the arcing. This is most likely due to the 20 eV metastable atoms ejecting secondary electrons from the surface. Those electrons are then accelerated in the large electric field colliding with the opposing wall. My solution was to mount the electrodes after the laser interaction where the atomic beam collimation is much better. A slit is also mounted directly in front of the electrodes to prevent metastable atoms from colliding with the electrode walls. The performance of the quenching electrodes is shown in Fig. 10. The separation of the electrodes is approximately 0.8 mm, and the operating voltage is 30 kV. The quenching efficiency is better than 99.9%.

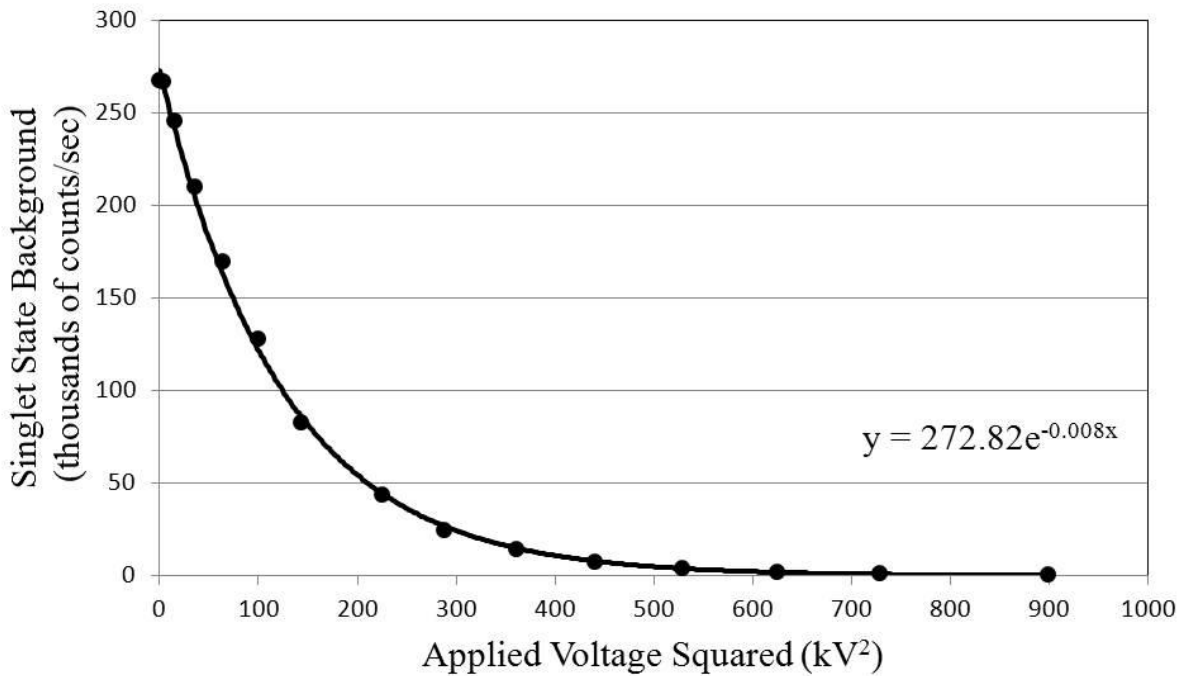

Fig. 10. Singlet state quenching performance.

## Signal Detection

When performing these experiments, it is necessary to have some method for detecting induced transitions. The signal could be derived from detecting photons by using a photo detector to measure laser absorption, or the detector could instead be used to measure photon re-emission. Alternatively, the metastable atoms themselves may be detected as long as the individual states may be distinguished from one another. The metastable triplet state of helium has three state, $m_s$ = +1, 0, and -1. The transitions used in this experiment depolarize the atom beam by having some probability of decaying into more than one state. This means that any one of those three states may be used to drive the transition; and to detect that the transition took place, a change of population is observed in any one of those three states. Preferable, the detection state is different from the interaction state to avoid looking at a small decrease in a large signal and is initially depopulated to have minimal background.

The three metastable triplet states are resolved by using Stern-Gerlach deflecting magnets to deflect the $m_s$ = ±1 states. The magnets deflect these states due to their non-zero magnetic dipole moment. The energies of the $m_s$ = ±1 states have a strong dependence on the magnitude of the magnetic field, which in a non-uniform field translates to a large energy dependence with respect to the position of the atoms. Anytime the energy of an object depends on the location of that object in space, it experiences a force

$$F = \frac{\partial E}{\partial x}.$$

This force acts in the direction that lowers the object's overall energy. The energy dependency of the metastable states is shown in Fig. 11 (represented as frequency). The $m_s$ = +1 states lower their energy at smaller magnetic fields (week field seekers). The $m_s = -1$ states lower their energy at larger magnetic fields (strong field seekers). In a magnetic field gradient, they deflect in opposite directions. So, given the magnetic field dependence of the atom's energy ($\partial E/\partial B$, $\pm 28$ GHz/T or $\pm 1.8 \times 10^{-23}$ J/T for the $m_s = \pm 1$ states) and a magnetic field gradient ($\partial B/\partial x$), the force the atom experiences is

$$F = \frac{\partial E}{\partial B} \cdot \frac{\partial B}{\partial x}.$$

By finding the component of the velocity perpendicular to the incident trajectory imparted by this force, the angle of deflection can be approximated by

$$\theta = \frac{Fl}{mv^2},$$

where $l$ is the length of the Stern-Gerlach magnet (3.8 cm), $m$ is the mass of the helium atom, and $v$ is the velocity of the atomic beam. In order to deflect the faster atoms ($\sim$2000 m/s) out of the atomic beam ($\sim$10 mrad), the gradient in the magnetic field must be at least 0.4 T/mm.

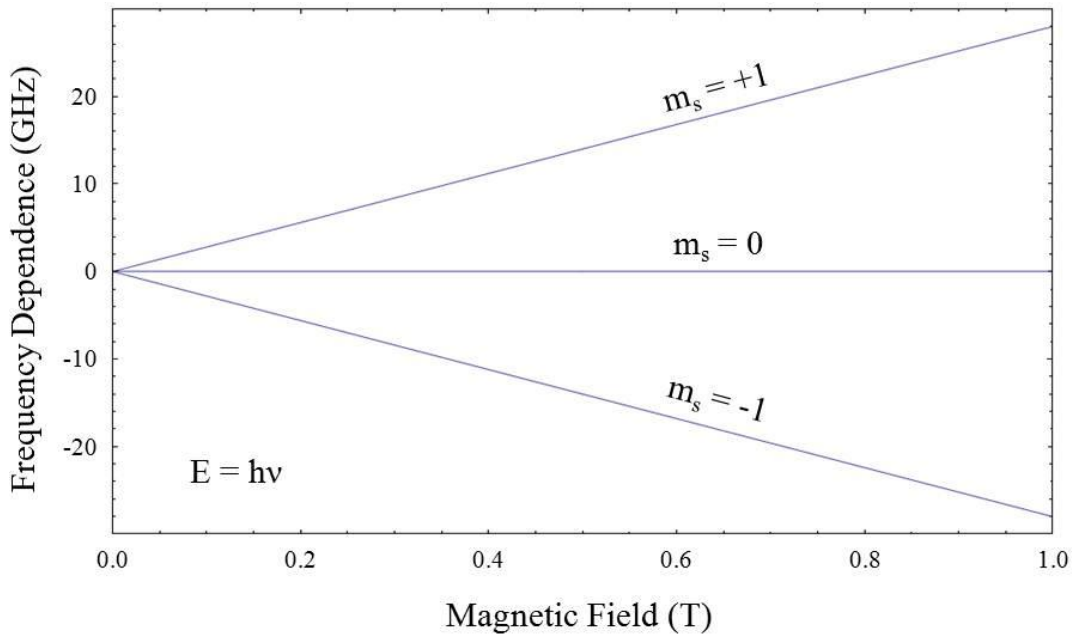

Fig. 11. Metastable energy magnetic field dependence.

One of the major changes from the old experimental setup to this new setup is which states are excited and which states are detected. Previously, the $m_s = 0$ state was used for the transitions and the $m_s = \pm 1$ states were detected. It was observed at the time that detecting deflected states systematically generated directional biasing in certain results. Using two oppositely deflected detection states created preferred beam directions that depended on how the transitions populated those states. In order to eliminate this effect, I designed this new apparatus to use the $m_s = \pm 1$ states for the interaction and to detect the $m_s = 0$ state. By using a single undeflected detection state for all transitions, there is no directional biasing caused by how the signal is detected. Unfortunately, detecting a straight-through beam is significantly more challenging due to all the potential sources of background. This includes the singlet states, the triplet $m_s = 0$ as well as the $m_s = \pm 1$ states, and high energy photons generated in the source. Many techniques have been incorporated into this experimental setup specifically to reduce the signal background. Background due to singlet and triplet $m_s = 0$ states is alleviated by the quenching electric field and the optical pumping which perform with better than 99.9% efficiency. Since the deflection angle for the $m_s = \pm 1$ states is strongly dependent on velocity as shown above, care must be taken that fast deflected atoms are not significantly contributing to the background. The dominant contribution to the background is actually the photons from the source. The atoms are detected with a channel electron multiplier. Though the efficiency of detecting photons should be much less than metastable atoms, the typical signal to noise ratio was about 20 to 1 (maximum signal). While the photons are still the dominant background, they are substantially reduced by detecting the atoms off a polished copper surface facing the electron multiplier (as shown in Fig. 12). The surface is at grazing incidence to reflect the photons away from the detector while allowing ejected electrons from the atoms to be pulled into the detector.

The copper surface is approximately 0.5 cm from the detector and is biased about 200 V more negative. The final maximum signal to noise ratio is around 200 to 1.

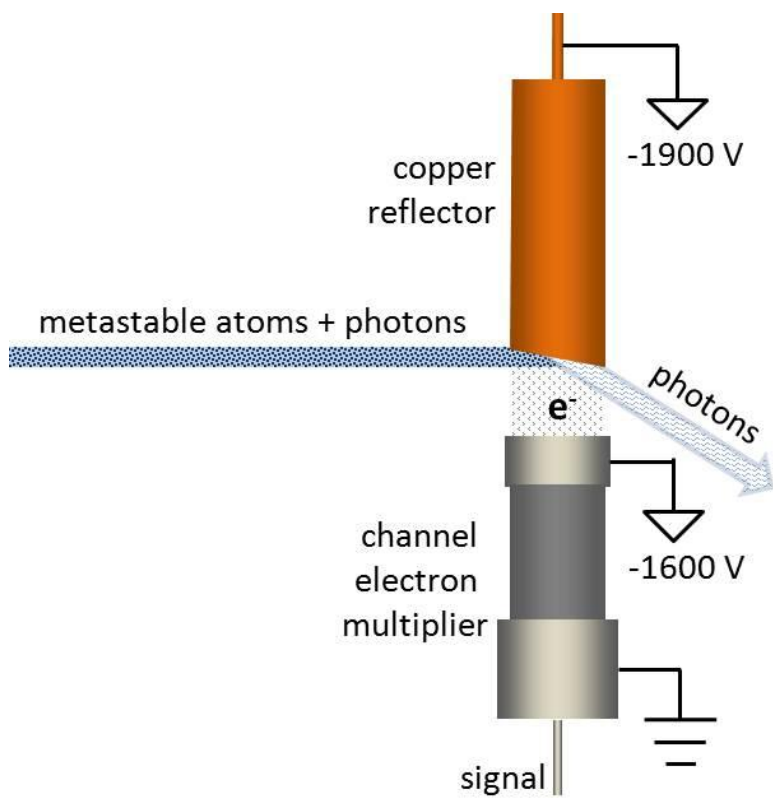

Fig. 12. Detector diagram.

# CHAPTER 4

## DATA ANALYSIS AND RESULTS

### Lineshape and Data Collection

Helium transitions from the 2S to 2P states ideally have a Lorentzian lineshape with a 1.6 MHz natural linewidth (FWHM). However, actual data collected on these transitions produce a slightly modified lineshape due to various broadening mechanisms. For an example of a lineshape measured in this experiment, Fig. 13 shows data collected on one of the transitions used to measure the J = 0 to J = 2 fine structure interval (J02; refer back to Fig. 1 in Chapter 1 for the notation used in this paper). The dashed line in the plot represents a simple Lorentzian distribution with a width equal to the natural linewidth and a matching height for comparison. The data were fit to a saturated Lorentzian fixed at the 1.6 MHz natural linewidth with various broadening mechanisms incorporated. These include Doppler related broadening due to velocity distribution and atomic beam collimation, as well as other effects like degree of saturation, Gaussian transit time broadening, and photon absorption recoil effects related to recycling atoms in the transition. The error bars for the residuals are one standard deviate and derived entirely from statistical $\sqrt{N}$ counting noise. The data for the fit were collected in about one hour and have an uncertainty in the line center of about 200 Hz. For the 20 data points, there were 10 fitted parameters with most being very close to their expected value. However, small adjustments in the values were necessary to achieve the quality of fit seen in the figure, with a final chi-squared of 34. It should be noted that no attempt was made to model drifts related to carrier frequency or alignment since it is difficult to know their exact relationship to the time the data was taken. Doubling the uncertainties brings chi-squared into good agreement with the degrees of freedom and accounts for any unknown fitting parameters like drifts.

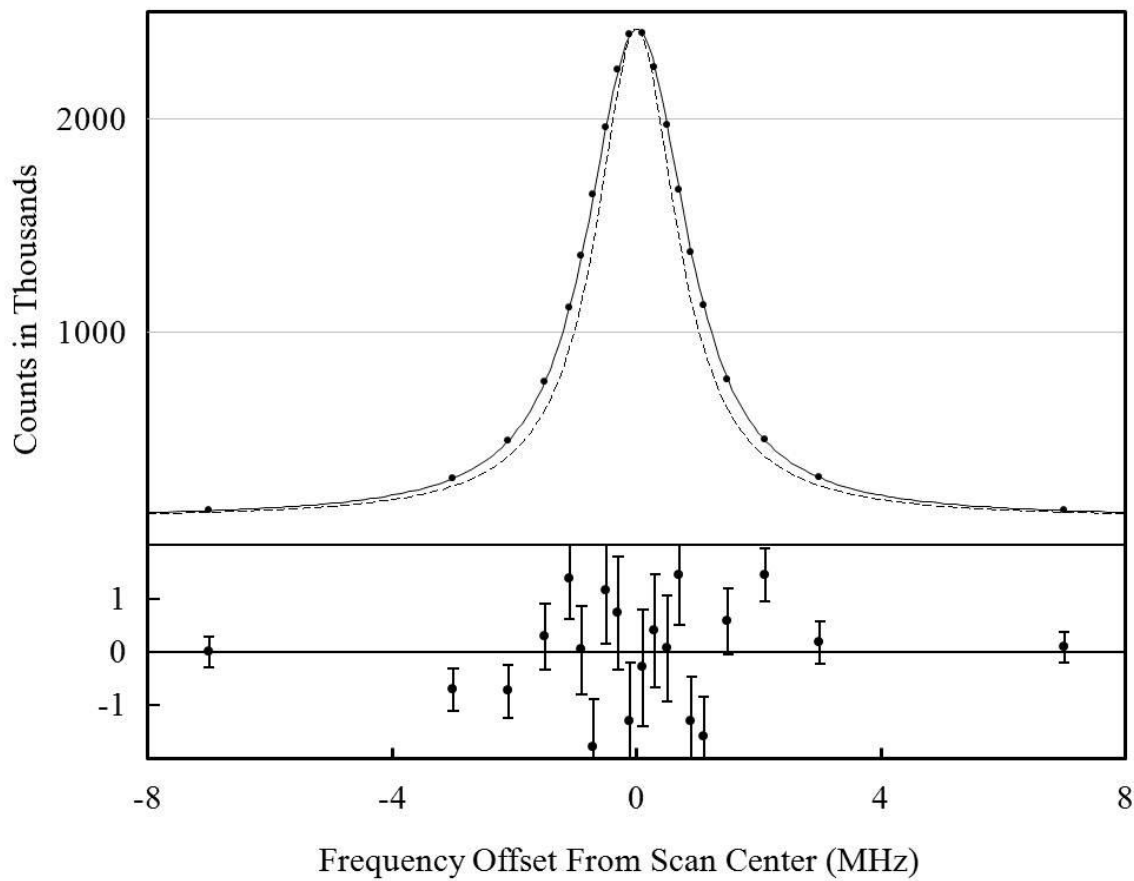

Fig. 13. Lineshape fit and residuals for the $2^3S_1$ ($m_s = +1$) to $2^3P_2$ ($m_j = 0$) transition.

While I do agree it is useful to perform fits like the one mentioned above since it helps for better understanding of many of the effects present in the data, measuring the final results in this way relies too heavily on the precise models used in those fits. Many times the parameters required in models are difficult to measure and could possible change from one data run to the next. Also, the fitting is very time consuming and requires a significant amount of computing power as the models become more and more complicated. I believe that time is better spent and the results are more reliable if the data are collected in such a way to eliminate the need for complicated models. For instance, most of the effects that broaden the lineshape do not affect the line center, i.e. velocity distribution, atomic beam collimation, and transit time broadening. Also, it reasonable to assume that effects which shift the line centers may, in fact, be subtracted out when calculating the actual intervals. This means that the only effects that do matter are those that shift one transition differently than another, and those that distort the lineshape. Unfortunately, the only way to produce a good fit is to model all of those effects and precisely determine the parameters used in those models.

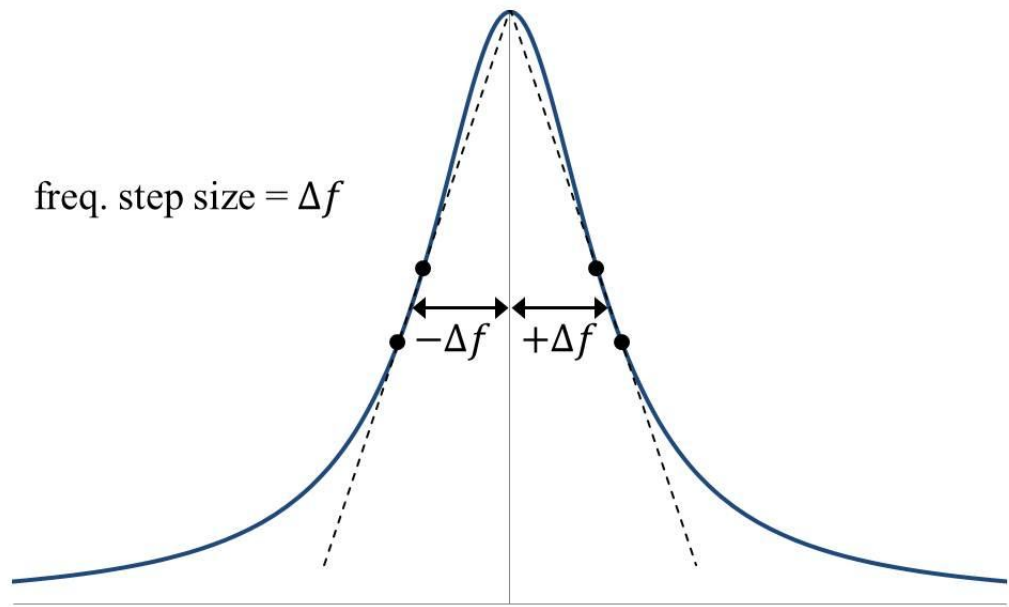

Fig. 14. Data collection method.

The data collection method used in this experiment is to measure the size of the signal and the slope on either side of the transition which is then used to calculate the line center. This is performed by collecting pairs of data points at fixed frequency step sizes from the predicted transition center (see Fig. 14). Running with different step sizes tests lineshape symmetry. Using this technique and concentrating only on the four levels used to measure the J02 interval, a typical data run takes about 2 hours to reach 300 Hz uncertainty. (Of course, this is the $\sqrt{N}$ statistical uncertainty and is simply provided as a reference when comparing to the final result, for which statistical uncertainty is negligible.) Long data runs are typically conducted overnight with statistical uncertainties around 100 Hz. Slow drifts in alignment, carrier frequency, or magnetic field are averaged out by stepping the frequency in a forward and reverse direction when collecting the data. Quickly measuring values with small statistical uncertainties makes possible more tests to look for systematic errors possibly due to a variety of factors. While many different factors have been investigated, notable systematic checks include tests related to power shift, magnetic field, Doppler alignment, and the previously mentioned frequency step size. Several consistency checks using other available $^4$He transitions as well as a very important calibration check using $^3$He test the reliability of the experimental setup and technique. All of these are discussed in detail in the sections that follow.

## Systematic Checks

The possibility of systematic errors is always important to be careful of when doing experimental physics, but it is especially so when doing high precision measurements. In fact, almost all the time spent collecting data is dedicated to conducting systematic checks. If not done carefully, systematic errors can be present when measurements are taken under a certain condition that in some way affects the data and therefore shifts the answer. If this condition is never varied, the answer may appear to be the correct but, in fact, is systematically off. The goal is to avoid having any significant systematic errors in the final result. Preferable this is done by eliminating the effect that is causing the systematic errors. However, some systematics cannot be eliminated and must therefore always be considered when conducting the experiment.

While systematics may be present in some measurements and not in others (I discuss some of these in the next section), of primary importance are systematic errors that affect the main result. For my dissertation, that result is the J = 0 to J = 2 fine structure interval in the $2\,^3P$ state of $^4$He. The four most important systematic checks examined on this interval are the power shift, frequency step size, magnetic field, and Doppler alignment. These are important either because a known systematic effect is present, or one may be present based on the way the data are collected and analyzed. All results discussed here are for the J02 interval and have been measured from both the $m_s = +1$ and $m_s = -1$ initial states.

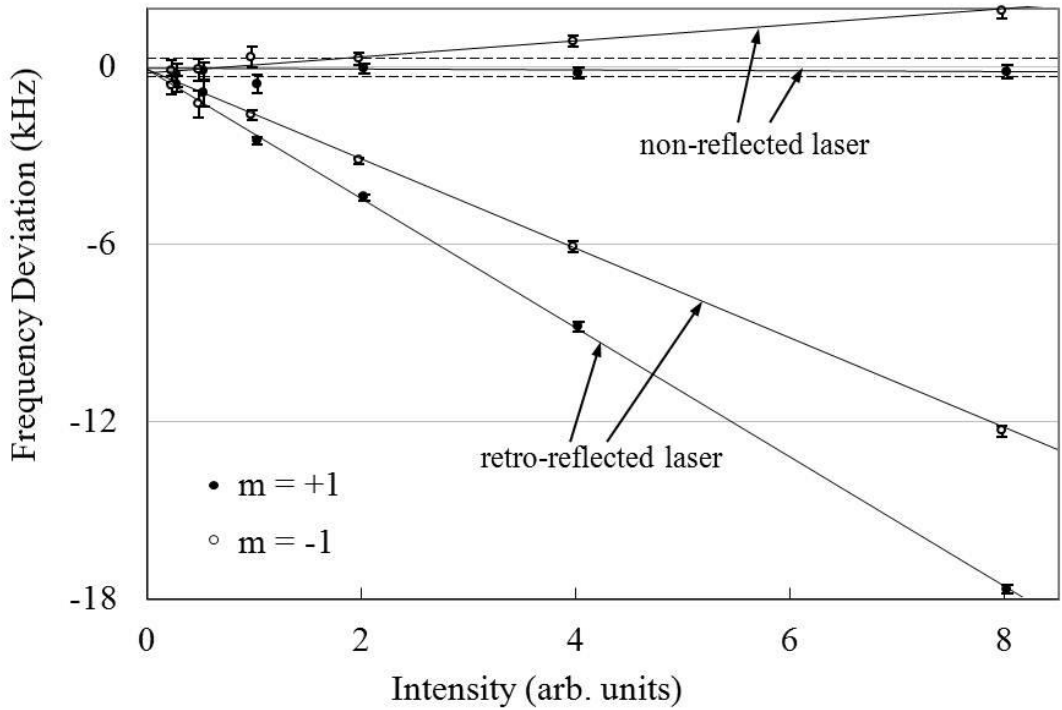

Fig. 15. Power shift systematic check.

The first systematic check to discuss is the power shift. The results for this test are shown in Fig. 15. A linear frequency shift is observed in the measured interval at higher laser intensities. This test was conducted using a retro-reflected interaction laser (2-way), which is used for Doppler cancellation, and a non-reflected interaction laser (1-way). All results have been extrapolated to zero power and are shown to be very linear to high intensity. As can be seen in the plot, the slope for the 2-way laser is much steeper than that of the 1-way laser. It should be noted, however, that the 2-way result is double the laser power of the 1-way result at the same laser intensity, since the laser beam interacts with the atoms twice. There are a few different effects present which I believe explain the slopes observed in the 1-way and 2-way results (discussed below). While it is useful to understand why the power shift occurs and how the slopes are affected, the power shift is not modeled in the data in order to determine the final result. All results are determined by collecting data at two laser intensities and extrapolating to zero power. The usual operating point for the intensities is chosen such that the final extrapolation is much less than 1 kHz.

I believe there are two major effects that contribute to the 1-way slope in the power shift. The first has to do with the velocity distribution of the atomic beam and the Doppler alignment. Slower atoms spend more time interacting with the laser and, therefore, are more likely to be excited. As the laser intensity increases, the slower atoms become saturated while the faster atoms contribute more to the signal. If the laser is perfectly aligned at a right angle to the atomic beam, there is no shift. However, as the laser becomes misaligned, the faster atoms experience a larger Doppler shift. This produces larger frequency shifts at higher laser intensities. Now of course, this would just be the slope that a single transition would experience. One might think that when calculating the intervals, this effect should cancel out. However, the transitions used to measure the intervals typically have different excitation rates, and therefore, they interact with the fast and the slow atoms differently. As is always the case, the slope in the interval is simply the difference in the slopes of the individual transitions. This effect is very alignment sensitive since misaligning the laser beam upstream and downstream yield opposite signs to the Doppler. This effect disappears as the laser approaches zero intensity because none of the atoms in the atomic beam are becoming saturated. The second effect on the 1-way slope is related to the recoil or pushing effect an atom experiences when absorbing a photon. All atoms recoil due to photon momentum when excited. The laser beam effectively pushes the atoms away, redirecting the beam that has undergone a transition. These atoms have some probability of decaying into the detection state as well as returning back to the state they came from (this, of course, is necessarily true since the rate up is always equal to the rate back down). The new population of atoms that decay back into the interaction state are now moving away from the laser which appears to be redshifted by one Doppler recoil (about 92 kHz for $^4$He absorbing a photon at 1083 nm). (As a side note, the atom recoils again when reemitting the photon during the decay, but that recoil averages over all directions and can be ignored.) After a single interaction, there is now a second atomic beam that has an increasingly large population at higher laser power and is more likely to undergo the interaction again. Also, as the original population starts saturating at higher power, this new population contributes more to the signal requiring a higher frequency to compensate for the redshift. This results in a positive slope to the power shift for the individual transitions when using the 1-way laser. This effect does go away at zero power since the probability of undergoing multiple interactions goes to zero.

The explanation for the slope observed in the 2-way laser is a little more complicated than that of the 1-way. The retro-reflected 2-way laser actually cancels out both of the effects described above for the 1-way laser since both of them are shifts related to the Doppler Effect. One would think that there shouldn't be a power shift for the 2-way or, at the very least, that it should be much smaller than that of the 1-way. Unfortunately, as can be seen in Fig. 15, the power shift is substantially larger. I believe this is due to a laser cooling effect. Laser cooling is a technique whereby two counter propagating lasers are used to cool down atoms or molecules by pushing the atoms with the photon momentum. This is done by driving the transitions on the low frequency side of the resonance curve. Atoms are more likely to interact with the blueshifted laser (i.e., the laser the atoms are heading into) since its frequency is shifted higher up the resonance curve. Just like the 1-way case above, after every interaction the atom is pushed away from the interacting laser, which is then redshifted by one Doppler recoil. Conversely, the counter propagating laser is blueshifted by the same amount. So, the interacting laser is moving down the resonance curve and the non-interacting is moving up the resonance curve. When the two lasers are at the same location on the resonance curve, the atoms are equally likely to interact with both and are kicked back and forth. This is the maximum laser cooling, within one Doppler recoil of zero velocity. For an atomic beam, laser cooling collimates the beam by reducing the transverse velocity, which increases the density with captured atoms that would have otherwise not passed through the detector's collimation slit. This argument does not apply when driving the transition on the high frequency side. With real laser cooling, very strongly recycling transitions are used to drive the transitions continually. For the effect observed in this experiment, it is limited by how likely the transitions are to decay back into the interaction state. So, when using the 2-way laser to determine the transition center, laser cooling increases the density of atoms, and thereby the signal, on the low frequency side. This is interpreted as the transition center being shifted lower in frequency. Since the effect is stronger at higher laser power where the atoms are more likely to undergo multiple interactions, the slope of the power shift is negative. This effect is very dependent on how strongly the transitions recycle. Once again, the slope of the interval is the difference in the slopes of the transitions. So just to be clear, all the transitions do have large negative slopes when using the 2-way laser, but the negative slope in Fig. 15 is only coincidental since the J = 0 transition is twice as likely to recycle as the J = 2 transition.

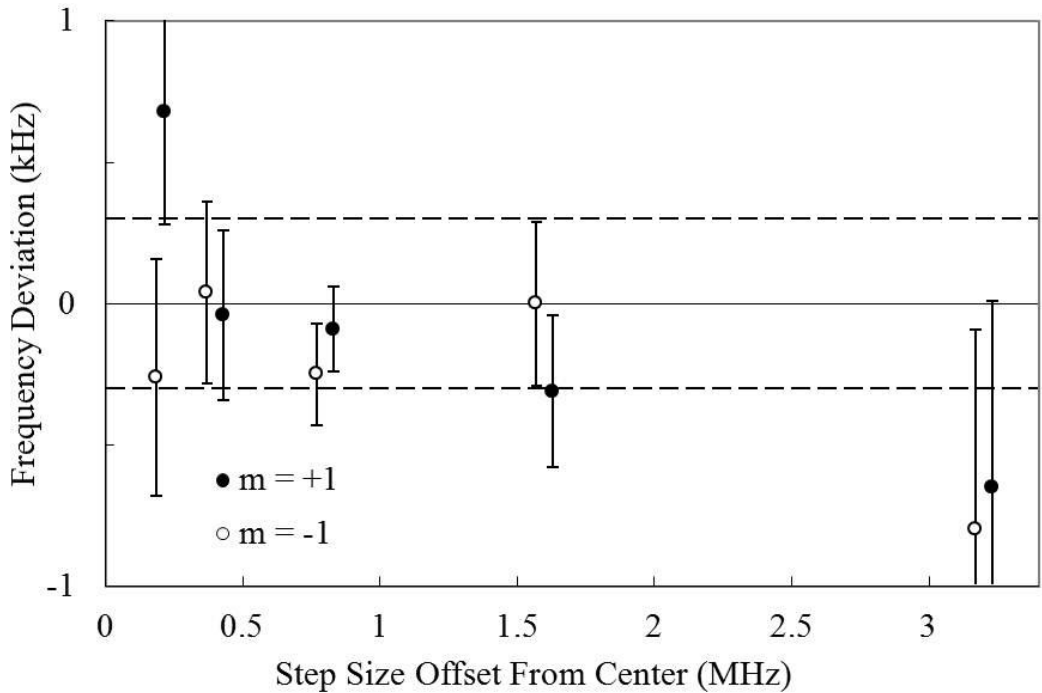

Fig. 16. Frequency step size systematic check.

The next systematic check is frequency step size. The results for this test are shown in Fig. 16. The plotted frequency deviation is with respect to the final quoted value for the J02 interval with the ±300 Hz uncertainty (dashed lines) for reference. This test is important since using a single frequency step size (refer back to Fig. 14 for how the data are collected) is not guaranteed to find the correct line centers if the distributions are not symmetric. That being said, it is actually the dependence of the measured intervals on step size that is of prime importance, not the transitions themselves. If both transitions being used to measure the interval have the same asymmetry, it will have no effect on the actual interval. For this test a range of step sizes was used from 0.2 MHz up to 3.2 MHz (the natural line width is ±0.8 MHz from the line center).

Collecting data and calculating the results at very large or very small step sizes is increasingly difficult due to the small slope of the transition and poor statistics. The standard operating position when collecting data is at 0.8 MHz where the statistics are good due to the steep slope and also the slope is relatively linear. Considering the statistical $\sqrt{N}$ uncertainty of the data, the results are consistent with the final quoted uncertainty.

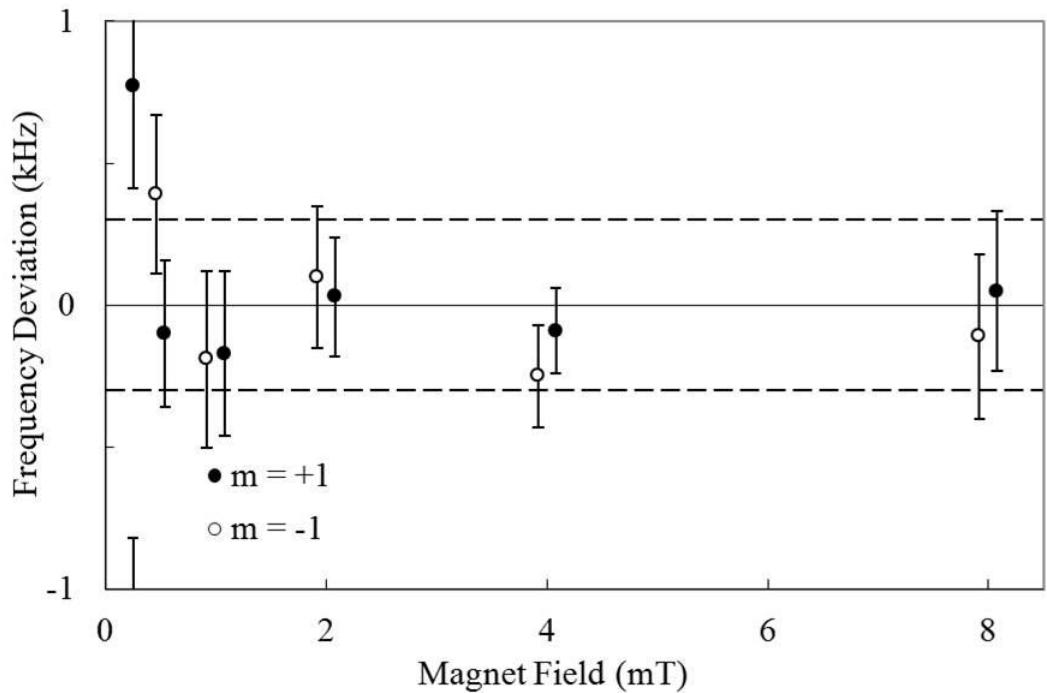

Fig. 17. Magnetic field systematic check.

The primary motivation behind the magnetic field systematic check is to test the calculated Zeeman corrections used to extrapolate the J02 interval to zero B-field (see Fig. 17).

Though the theory behind these corrections is very well understood, the test serves to verify the

implementation used in this experiment. Measuring consistent results over a large range of B-

field values demonstrates the theory is being properly applied and the calculated corrections are evaluated to the necessary precision.  For this systematic check, values ranging from 0.25 mT up to 8 mT were used.  This corresponds to actual Zeeman level shifts from approximately 7 MHz up to 224 MHz (~28 MHz/mT) for the transitions used to measure the J02 interval.  Though these require very large corrections, the transitions used to determine the intervals have very similar dependencies on magnetic field.  In fact, the large B-field dependence comes from $m_s = \pm 1$ metastable states which is entirely linear since these are pure states (this completely subtracts out since the interval is always measured using the same initial state).  The small residual dependence comes from the non-pure $m_j = 0$ upper states.  States with the same $m_j$ value couple with one another and, therefore, are not pure.  The tendency is for the states to repel each other at higher B-field.  This has primarily a quadratic dependence on the magnetic field equal to 0.033 MHz/mT$^2$, which amounts to 2.1 MHz at 8 mT.  The magnetic field systematic check is also useful for analyzing the effect of overlapping transitions.  At higher magnetic field values, the transitions are well separated, but as the magnetic field is lowered, nearby transitions begin to overlap.   As shown in Fig. 17, the results at 0.25 mT are no longer consistent within the 300 Hz uncertainty due to overlapping transitions.

The last of the four primary systematic checks mentioned above is Doppler alignment (see Fig. 18).  This test examines the dependence of the J02 interval on the 1$^{st}$ order Doppler effect.  The 1$^{st}$ order Doppler effect is zero when the laser beam is exactly perpendicular to the atomic beam.  With an average velocity of 1600 m/s, the maximum Doppler shift for the atomic beam is around 1.5 GHz.  This test was evaluated with Doppler misalignments between $\pm 600$ kHz, which corresponds to angular misalignments between $\pm 400$ $\mu$rad from perpendicular.  During normal operation, the laser is easily Doppler aligned to within $\pm 10$ kHz to the atomic beam.  For determining the J02 interval, this test is very sensitive to the average direction and velocity of the atoms that the individual transitions sample. These results indicate that systematic effects due to small misalignments are negligible.

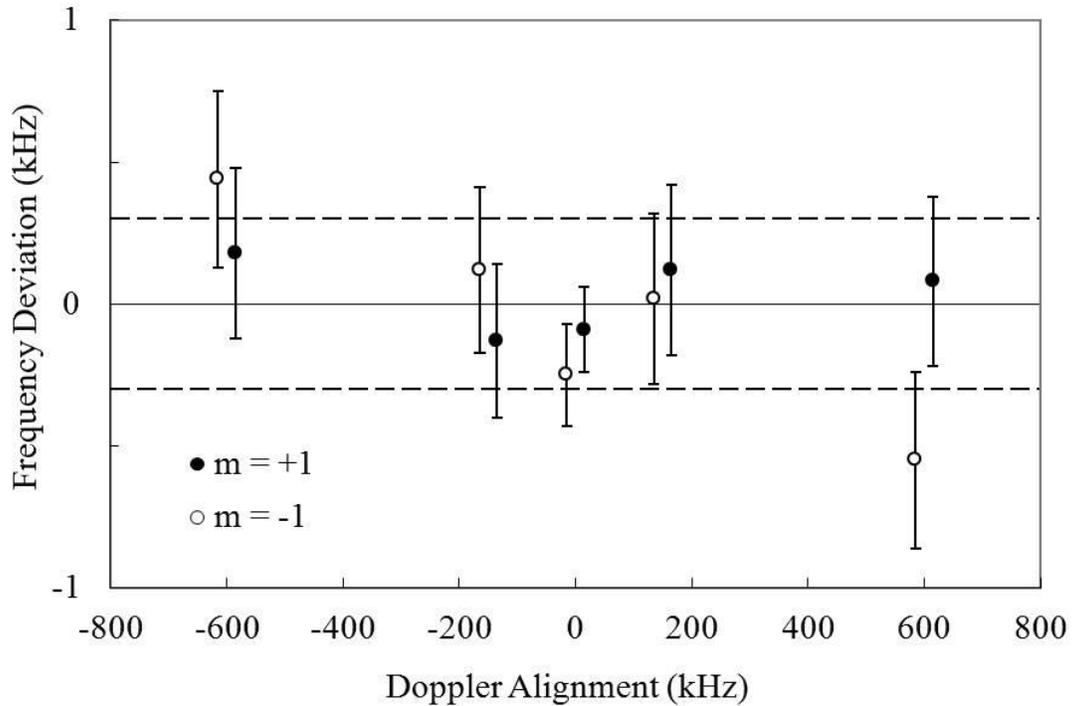

Fig. 18. Doppler alignment systematic check.

The systematics discussed above are certainly important to study and understand. However, the systematics that remain unknown in the setup are, for obvious reasons, the most challenging to deal with. Numerous parameters and conditions on the experiment have been adjusted to hunt down hidden systematic effects, e.g., changing the helium pressure, rotating the laser polarization, rotating the magnetic field, creating a non-uniform magnetic field, changing the singlet state quenching electric field, misaligning the detection slit, changing the EOM modulation power, moving the carrier frequency, etc. Ultimately, it comes down to turning every knob on the experiment until there are no more knobs to turn, metaphorically speaking.

None of these other tests had any significant impact of the result for the J02 interval. The final uncertainty budget for this interval is given in Table 2.

Table 2. Uncertainty budget for the J = 0 to J = 2 fine structure interval.

| Source | Uncertainty (kHz, one standard deviation) |
| --- | --- |
| Laser Power | < 0.1 |
| 1st Order Doppler | < 0.1 |
| B-field | < 0.1 |
| Lineshape | 0.2 |
| Other | 0.1 |
| Total (quadrature sum) | 0.3 |

## Consistency Checks

A number of consistency checks have been performed to test the reliability of the experimental setup and technique. These include several tests using various intervals in $^4$He, as well as one extremely valuable measurement done using $^3$He that serves as a type of calibration to the experiment. Though these tests do not necessarily incorporate the same systematics present in the primary fine structure interval that is the focus of this dissertation, they do verify that the experiment is working properly and is capable of yielding reliable measurements at the level of the quoted uncertainty. Therefore, the consistency checks discussed below substantially increase the confidence in the final results.

In this experimental setup, there are a total of eight available $^4$He transitions when exciting from the $m_s = \pm1$ metastable states. The maximum number of independent intervals between those transitions is, therefore, seven (refer back to Fig. 1 in Chapter 1 for an illustration of the chosen intervals used here). Two of those are the fine structure intervals, the J = 0 to J = 2 interval (J02) and the J = 1 to J = 2 interval (J12). Alternatively, the J = 0 to J = 1 interval could be used as one of the two independent fine structure intervals. However, for this setup, it is not preferred since the transitions used to measure that interval have very different B-field dependencies and require orthogonal laser polarizations to interact. A third independent interval must be used to measure the magnetic field. This is done by measuring one or more of the Zeeman intervals within each of the J states. Those are the minimum three intervals required to measure the fine structure in $^4$He. The last four independent intervals are all consistency checks (shown in Fig. 19; lightly shaded results use the secondary scale on the right). The first two consistency checks compare results for the J02 and J12 fine structure intervals when independently measured from the $m_s = +1$ and $m_s = -1$ initial states. These are very consistent within the quoted 300 Hz uncertainty. The last two are both tests that use Zeeman intervals, which include the metastable interval between the $m_s = \pm1$ states (S1m+1/-1) and a comparison between the two intervals in the J = 2 state (J2m0/+1 and J2m0/-1). Both of the tests involving Zeeman intervals are consistent when they are equal to zero, after applying the magnetic field corrections determined by an independently measured B-field.

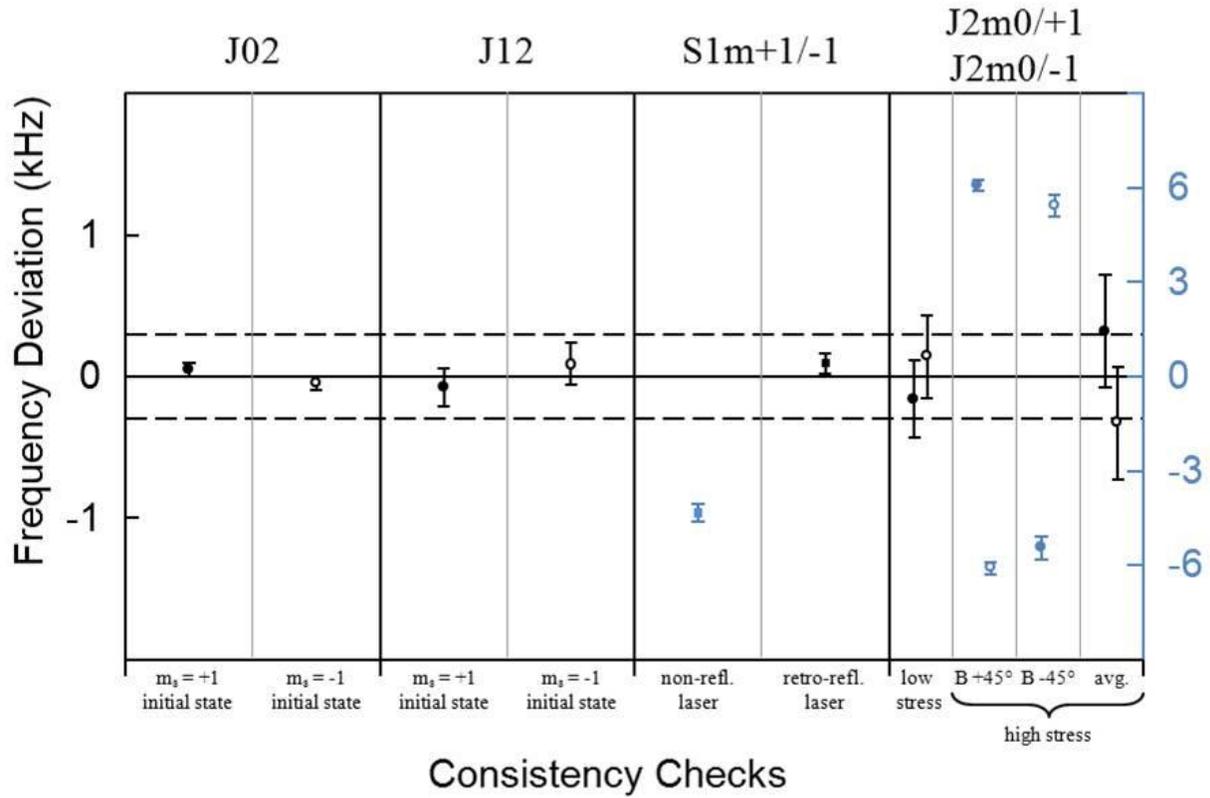

Fig. 19. Consistency check results.

The S1m+1/-1 test is clearly inconsistent when measured with the non-reflected interaction laser. This is because the $m_s = \pm1$ metastable states are effectively two independent atomic beams, which are not guaranteed to have the same average direction. So, in fact, comparing the two states reveals that a residual Doppler of several kilohertz systematically affects the results, which the magnetic field corrections cannot account for. Using the retro-reflected laser to measure the S1m+1/-1 interval resolves this inconsistency by canceling the residual Doppler shift. The last consistency check, comparing J2m0/+1 and J2m0/-1, also shows an inconsistency due to a systematic effect. This test is sensitive to the polarization of the interaction laser. The transitions used to measure the J2m0/+1 and J2m0/-1 intervals require orthogonal polarizations with respect to the quantization axis (i.e. magnetic field). Transitions that change the m value ($m_s = \pm 1$ to $m_j = 0$) are driven by a perpendicular laser polarization to the magnetic field direction, and conversely, those that do not change the m value ($m_s = \pm 1$ to $m_j = \pm 1$) are driven by a parallel laser polarization. It has been observed that the two orthogonal laser polarizations for the interaction laser do not in general have the same average direction. When this happens, the comparison of the B-field corrected results exhibits a residual Doppler shift. The two intervals shift in opposite directions since this test is essentially a comparison of the $m_j = 0$ position with the average position of the $m_j = \pm 1$ states. So, for instance, as J2m0/+1 interval grows bigger, the J2m0/-1 becomes smaller by the same amount. This polarization effect is created by stress-induced birefringence in the laser port window on the apparatus which, on average, redirects the polarization states of the laser. This effect can be reduced by carefully eliminating stresses on the window, e.g. loosening the clamp that holds the window on the apparatus. As a further test on the polarization hypothesis, I rotated the magnetic field by 90 degrees, switching the laser polarization states the transitions sample. As expected, this flipped the sign of the effect, and averaging the effect over the polarization yields a result consistent with zero. It is important to note that the polarization and initial state effects observed in these last two consistency checks have no systematic influence on the J02 fine structure interval, nor would they be expected to. This is because the transitions in J02 interval are measured from the same initial state and use the same laser polarization.

The very precisely know hyperfine structure interval in the metastable state of $^3$He serves as an extremely valuable consistency check for this experiment. The value measured by Rosner and Pipkin has an uncertainty of only 16 Hz [24], which is almost 20 times smaller than the uncertainty quoted for the J02 interval in this work. So unlike $^4$He, this offers the unique opportunity to measure a large interval (6.7 GHz) and, essentially, compare the result to the correct answer. The experimental setup preceding mine was never able to measure successfully this hyperfine structure interval consistent to better than 1 kHz [12]. The limitations to the previous experimental setup have been resolved in my new setup with the implementation of two major modifications: optical pumping to prepare the initial state of the atomic beam, and the use of a single undeflected detection state. The previous experimental setup interacted with the "$m_s = 0$" states [$(F = \frac{1}{2}, m_f = -\frac{1}{2})$, $(F = \frac{3}{2}, m_f = +\frac{1}{2})$] and detected the "$m_s = \pm 1$" states [$(F = \frac{1}{2}, m_f = +\frac{1}{2})$, $(F = \frac{3}{2}, m_f = +\frac{3}{2})$ and $(F = \frac{3}{2}, m_f = -\frac{1}{2})$, $(F = \frac{3}{2}, m_f = -\frac{3}{2})$] by deflecting them around a stopwire (the six $m_f$ states in the metastable state of $^3$He resemble pure $m_s$ states at high magnetic field). A Stern-Gerlach deflecting magnet was used to prepare the initial state of the atomic beam. However, this created a large Doppler shift of several hundred kilohertz between the metastable hyperfine levels and required extremely reliable Doppler cancellation to remove. Also, the detection of multiple deflected states caused a dependence on stopwire position due to directional biasing. The new experimental setup has been designed and tested to eliminate both of these effects. The new results for the $^3$He hyperfine structure interval are consistent with the measurement by Rosner and Pipkin to about 100 Hz with an uncertainty of 80 Hz. The measurements were done with standard operating conditions with no systematics investigated, and the uncertainties are due almost entirely to statistical $\sqrt{N}$ counting noise. A comparison of the results is shown in Fig. 20. The quoted 300 Hz uncertainty for the J02 interval is displayed for reference.

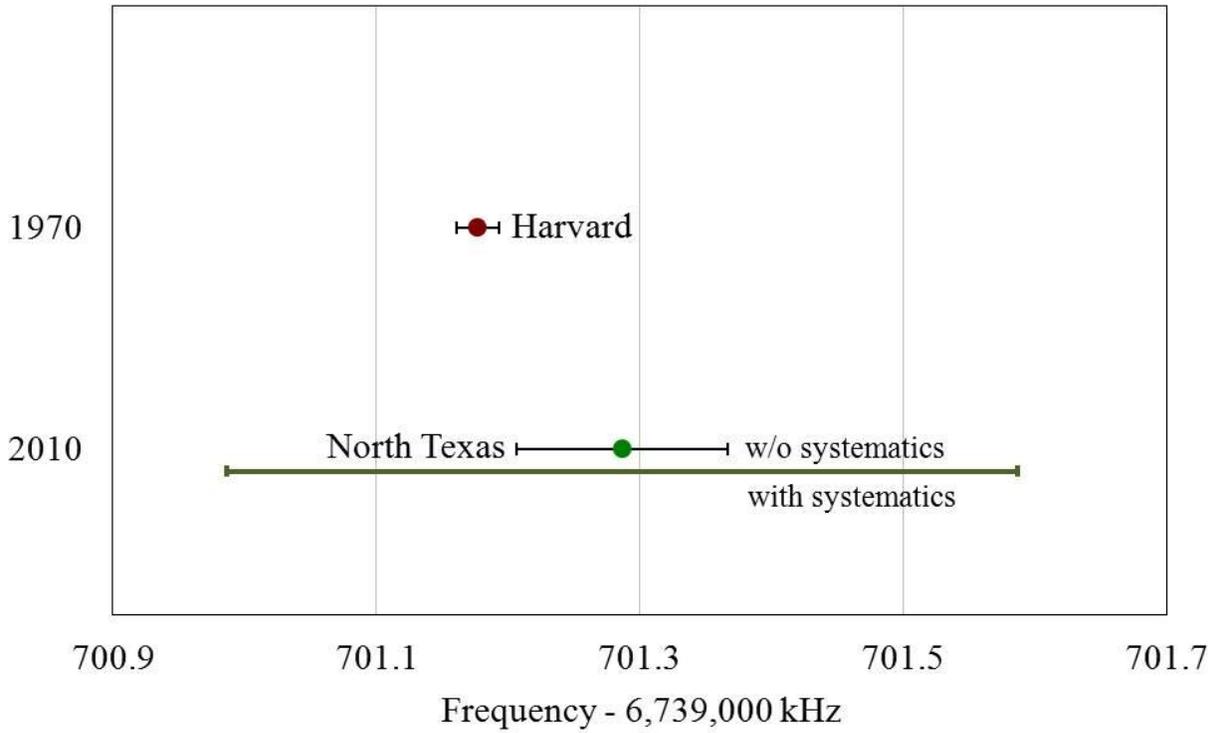

Fig. 20. $^3$He hyperfine interval result comparison.

Results and Comparison

The final result for the J = 0 to J = 2 fine structure interval (J02) is 31,908,131.25 kHz

with an uncertainty of ±300 Hz. This result along with a previously reported value for the J = 1

to J = 2 interval (J12) (measured with the old experimental setup) and an updated value for the

J = 0 to J = 1 interval (J01) are given in Table 3. The updated value for the J01 interval is calculated by taking the difference of the J02 and J12 intervals. Its uncertainty is the uncertainties of the other two values added in quadrature.

Table 3. [4]He fine structure interval results.

| [4]He Fine Structure Interval | Measured Value (kHz) |
|---|---|
| J = 0 to J = 2 | 31,908,131.25 (0.30) |
| J = 1 to J = 2 | 2,291,175.9 (1.0)  [12] |
| J = 0 to J = 1 (J02 – J12) | 29,646,955.35 (1.04) |

A comparison of measured values for each of the three [4]He fine structure intervals with quoted uncertainties of 3 kHz or less is shown by year reported in Fig. 21. A very recently reported theoretical calculation for the intervals is displayed as well. Since the majority of groups report on the J01 rather than the J02 interval, the updated J01 interval is displayed for comparison. My advisor, Dr. David Shiner, previously worked on the results at Yale University and subsequently continued that work at the present location, the University of North Texas. Therefore, the Yale results were measured with a similar technique. For this reason, my new value for the J02 interval can be thought of as an updated result to that older value (though with obviously significant changes to the experimental setup and many of the techniques). The Harvard group is the only other group actively reporting on the J02 interval. My new value clearly does not agree with the value they reported in 2005. The difference between the two values is about 4.5 kHz. With the Harvard group's 0.94 kHz quoted uncertainty, the combined effect corresponds to 4.5 standard deviations.

What turns out to be a very interesting comparison is my new J02 value with the recently published theoretical calculations for this interval. For a long while now, theory has not agreed well with experiment, being off by tens of kilohertz. In 2009, Pachucki and Yerokhin published new theoretical calculations for each the $^4$He fine structure intervals complete for all terms up to $m\alpha^7$ [25]. However, due to an error in the calculations, the results for the intervals were off. Namely, the J02 interval was reported to be nearly 8 kHz smaller than what is shown in Fig. 21. Remarkably, a publication in Physical Review Letters [9] and, shortly afterwards, an erratum for the previous article [26] was published earlier this year, in 2010, with corrected values for the fine structure intervals, which are shown in the figure. The updated value for the J02 interval is almost directly on top of the value reported in this dissertation. My experimental value and the theory are in agreement to within 60 Hz. The 1.2 kHz uncertainty of the J02 theoretical calculations was reported as an estimate of uncalculated higher order terms [25]. In fact, the calculated terms are reported to have an uncertainty on the order of 20 Hz [9]. Considering the new measurement for J02 presented here and the close agreement with the calculations, this could suggest that 1.2 kHz is an over estimate, and the higher order terms are much smaller. In fact, there was already an estimate of the leading $m\alpha^8 \ln\alpha$ terms which was reported to be only 300 Hz [27]. Also, when reporting the uncertainties for the uncalculated terms, the J02 uncertainties were estimated to be smaller than for the other two fine structure intervals. The reason given was that the J02 interval is independent of singlet-triplet mixing terms that are present when evaluating intervals that involve the J = 1 level. This could also suggest that there are advantages in using the J02 interval to compare to theory.

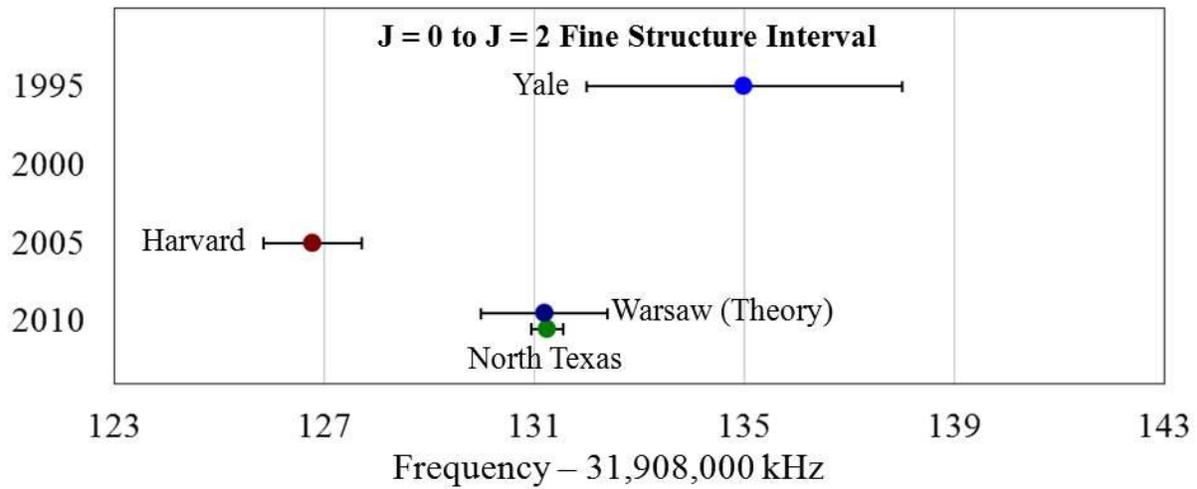

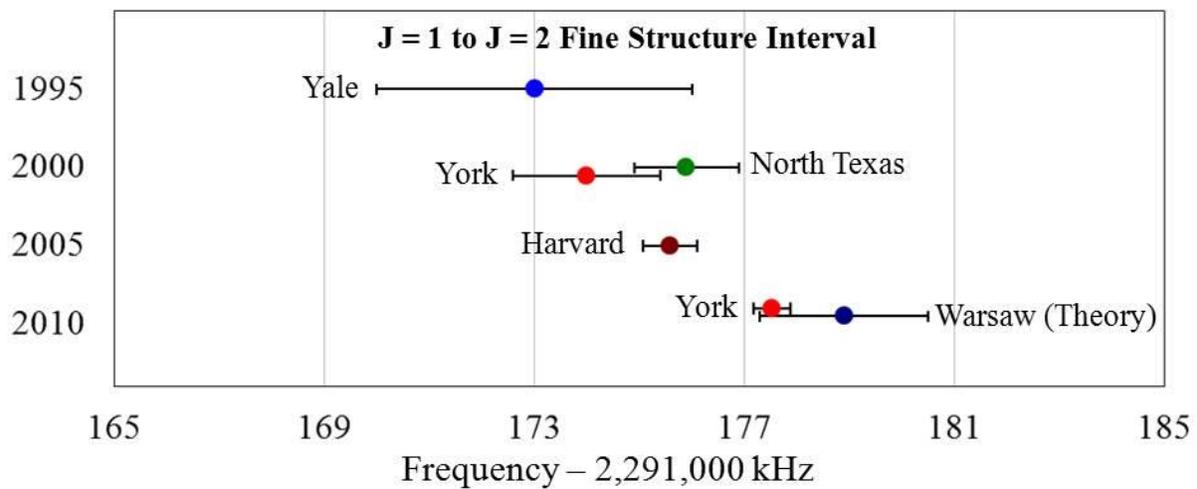

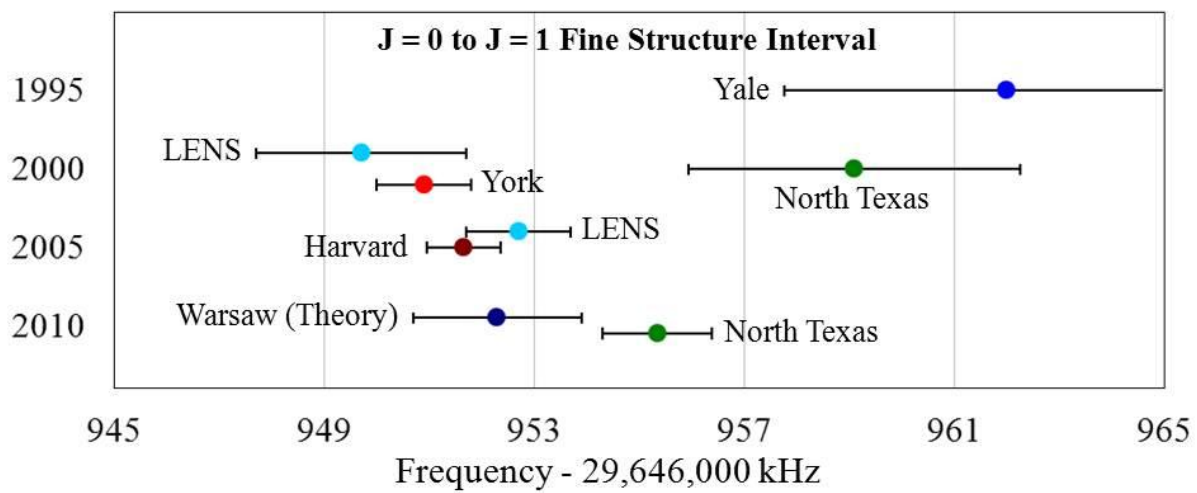

Fig. 21. Comparison of fine structure intervals by year.

The agreement shown here between experimental measurements and the more complete theoretical calculations certainly tests QED and helium theory at a new level of precision, but interestingly, for the first time in recent history it displays encouraging evidence that helium fine structure may offer a competitive alternative to the determination of the fine structure constant $\alpha$. By far the most precise determination of $\alpha$ comes from measurement of the electron magnetic moment or g-factor [1], an order of magnitude better than any competing method. However, without an independent alternative measurement of $\alpha$, there can be no test on theory. The best alternatives to date come from Rb [28] and Cs [29] recoil measurements. Now, if we consider only the uncertainty in the calculated terms for the theory along with the experimental uncertainty measured here (which for helium fine structure measurements is 2.5 times more sensitive in determining $\alpha$ than any other reported uncertainties to date), the helium fine structure offers a competitive value for a new alternative determination of the fine structure constant. By comparing the $J = 0$ to $J = 2$ fine structure interval presented in this dissertation to the latest theoretical calculation [9], a new value for the fine structure constant has been evaluated to yield the value $\alpha^{-1} = 137.035\ 999\ 55(64)$. This new value is presented in Fig. 22 along with the other best determinations of $\alpha$ to date.

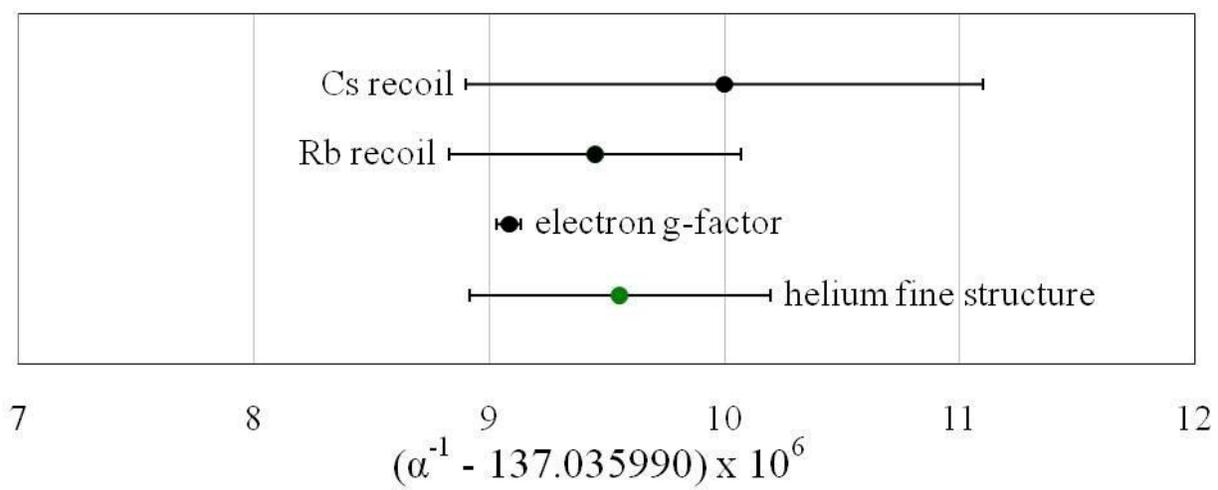

Fig. 22. Fine structure constant comparison.

CHAPTER 5

CONCLUSION AND REMARKS

For my dissertation, I have successfully designed and built an experimental apparatus that I have used to measure the J = 0 to J = 2 fine structure interval in the $2^3P$ state of $^4$He. The final result for this measurement is 31,908,131.25 kHz with an uncertainty of ±300 Hz. At better than 1 part in $10^8$, this value is 3 times more precise than any other reported measurement for this fine structure interval. There is a discrepancy of 4.5 standard deviations with another precision measurement of the J = 0 to J = 2 interval, which was reported in 2005 by the group at Harvard. However, a very recent theoretical calculation by Pachucki and Yerokhin (with only 20 Hz uncertainty in the calculated terms) is in near perfect agreement with the value I have measured. The difference between theory and this experiment is only 60 Hz, corresponding to only 0.2 standard deviations. Additionally, this recent theoretical development along with my measured value for this fine structure interval has now allowed for a new determination for the fine structure constant, yielding $\alpha^{-1}$ = 137.035 999 55(64). This is in good agreement with the best determination of $\alpha$ using a measurement of the electron g-factor and is competitive with the best alternative methods involving Rb and Cs recoil measurements.

The results obtained for this dissertation were measured using many of the techniques of the preexisting experimental setup but with an entirely rebuilt apparatus. Due to certain limitations inherent to the old setup, key changes along with major improvements were incorporated into the new experimental setup. Most notably, the interaction state used for this experiment is now the metastable $m_s = ±1$ states, rather than $m_s = 0$. This not only allowed for more consistency checks than were previously available (e.g., obtaining two independent measurements for the J = 0 to J = 2 interval), but also eliminated a systematic bias (related to how the atoms were detected) observed in certain key measurements. As a substantial portion of my research, I built a new apparatus incorporating key technologies designed for the purpose of conducting the experiment in this way. Some of the major improvements incorporated into the new setup include designing and building a custom fiber laser for optical pumping, high voltage electrodes for metastable singlet sate quenching, and a greatly improved metastable source which has allowed for more reliable operation and much larger signals for better statistics.

With more precise measurements for the helium fine structure intervals and the progress made in the theoretical calculations, more rigorous tests are made possible on helium theory and the techniques used in QED calculations. With the techniques used in this experiment and the work that I have accomplished to push those techniques to new levels of precision, I believe that continued progress can be made. One improvement would be to stabilizing the laser intensity of the individual sidebands used to drive the transitions, as opposed to the entire laser intensity. Though the step in frequency is very small from one side of the transition to the other, this would ensure that no small changes in laser intensity are systematically biasing the result. A possible method to accomplish this is to use narrow bandwidth (3 to 6 GHz) Bragg gratings in the fiber to selectively reflect the sideband driving the transition. A beam splitter can be used to send the reflect sideband to the apparatus. Another limitation for obtaining precision results well beyond the 300 Hz level presented here is the time required to reach the necessary statistical uncertainties when collecting data. Increasing the signal size would reduce that time and improve the statistics. It is possible that this could be done by incorporating transverse laser cooling on the atomic beam, increasing the beams density as well as improving the collimation of the beam by as much as a factor of 5. This cooling could also eliminate the large power shift observed in the retro-reflected laser beam data. This is because the atomic beam would already be cooled to within one Doppler recoil of perfect collimation and would no longer be sensitive to further cooling effects caused by the interaction laser.

APPENDIX

HELIUM TRANSITIONS USING MATHEMATICA

This appendix contains the full Mathematica code used to calculate the transition energies and transition rates for $^4$He. This is used to not only predict the relative transition energy locations and expected signal sizes, but the very important Zeeman corrections evaluated and fit to polynomials, which are used in the actual data analysis for the experiment.

Load Packages and Definitions and error message preferences

```
<< "LinearAlgebra`";
Off[ClebschGordan::"phy"]
Off[Part::"pspec"]
<< "VectorAnalysis`"
Needs["EDA`Master`"]
Off[General::"spell"] Off[General::"spell1"]
```

Convert Cartesian vectors into irreducible vectors

```
ir[v_] := { (v[[1]] - i v[[2]])/√2 , v[[3]], - (v[[1]] + i v[[2]])/√2 };
```

Convert irreducible vectors into Cartesian vectors

```
vec[v_] := Simplify[{ (v[[1]] - v[[3]])/√2 , (i (v[[1]] + v[[3]]))/√2 , v[[2]] }];
```

Lorentzian distribution function

```
lorz[int_, fr_, f_] := int / (1 + ((f - fr) 1/0.8)^2) ;
```

Define operators for tensor direct product, complex conjugate, vector dot product and triple vector dot product

```
x_ ⊗ y_ := BlockMatrix(Outer[Times, x, y]);

x_* := Conjugate[x];
x_ · y_ := x.y;
x_ · y_ · z_ := x.y.z;
```

Returns eigenvalues and normalized eigenvectors sorted according to energy from low to high

$$\textbf{EigenSys[op\_]} := \text{Module}\Big[\{\text{val}, \text{vect}, \text{sval}, \text{svect}\},$$

$$\{\text{val}, \text{vect}\} = \text{Eigensystem}[\text{op}];$$

$$\text{vect} = \text{Table}\Big[\frac{\text{vect}[[i]]}{\sqrt{\text{vect}[[i]].\text{vect}[[i]]}}, \{i, 1, \text{Length}[\text{vect}]\}\Big];$$

$$\text{sval} = \text{Sort}[\text{val}]; \text{svect} = \{\};$$

$$\text{Do}[\text{If}[\text{sval}[[i]] == \text{val}[[j]], \text{svect} = \text{Append}[\text{svect}, \text{vect}[[j]]]], \{i, 1, \text{Length}[\text{val}]\}, \{j, 1, \text{Length}[\text{val}]\}];$$

$$\{\text{sval}, \text{svect}\}\Big];$$

Define magnetic field values to evaluate

$$\textbf{bvalue} = 40;$$

$$\textbf{brange} = \text{Table}[b \rightarrow i, \{i, 0, 100, 1\}];$$

Input basic operator building blocks

Calculate electric dipole matrices corresponding to x - iy, z, and x + iy.

$$\textbf{ed} = \text{Simplify}\Big[\text{Table}\Big[$$

$$\int_0^\pi \int_0^{2\pi} Y_{lf}^{mf}(\theta, -\phi) \sqrt{4\pi} \; Y_1^q(\theta, \phi) Y_{li}^{mi}(\theta, \phi) \sin(\theta) \, d\phi \, d\theta, \{q, -1, 1\}, \{lf, 0, 1\}, \{mf, -lf, lf\},$$

$$\{li, 0, 1\}, \{mi, -li, li\}\Big]\Big];$$

$$\textbf{edm} = \text{Table}[\text{Flatten}[\textbf{ed}[[i]]], \{i, 1, 3\}];$$

$$\textbf{Sq} = \sqrt{\text{Length}[\textbf{edm}[[3]]]} \;;$$

$$\textbf{edm} = \text{Table}[\text{Table}[\textbf{edm}[[i]][[r + (q-1)\,\textbf{Sq}]], \{q, 1, \textbf{Sq}\}, \{r, 1, \textbf{Sq}\}], \{i, 1, 3\}];$$

Table[MatrixForm[**edm**[[i]]], {i, 1, 3}]

$$\left\{ \begin{pmatrix} 0 & 0 & 0 & -1 \\ 1 & 0 & 0 & 0 \\ 0 & 0 & 0 & 0 \\ 0 & 0 & 0 & 0 \end{pmatrix}, \begin{pmatrix} 0 & 0 & 1 & 0 \\ 0 & 0 & 0 & 0 \\ 1 & 0 & 0 & 0 \\ 0 & 0 & 0 & 0 \end{pmatrix}, \begin{pmatrix} 0 & -1 & 0 & 0 \\ 0 & 0 & 0 & 0 \\ 0 & 0 & 0 & 0 \\ 1 & 0 & 0 & 0 \end{pmatrix} \right\}$$

Calculate orbital angular momentum tensor.

$$\textbf{Ltensor} = \text{Simplify}\Big[\text{Table}\Big[$$

$$\int_0^\pi \int_0^{2\pi} Y_1^{mf}(\theta, -\phi)\sqrt{\frac{4\pi}{5}}\; Y_2^q(\theta, \phi)\, Y_1^{mi}(\theta, \phi)\sin(\theta)\,d\phi\,d\theta, \{q, -2, 2\}, \{mf, -1, 1\}, \{mi, -1, 1\}\Big]\Big];$$

Table[MatrixForm[**Ltensor**[[i]]], {i, 1, 5}]

$$\left\{\begin{pmatrix} 0 & 0 & -\frac{\sqrt{6}}{5} \\ 0 & 0 & 0 \\ 0 & 0 & 0 \end{pmatrix}, \begin{pmatrix} 0 & \frac{\sqrt{3}}{5} & 0 \\ 0 & 0 & -\frac{\sqrt{3}}{5} \\ 0 & 0 & 0 \end{pmatrix}, \begin{pmatrix} -\frac{1}{5} & 0 & 0 \\ 0 & \frac{2}{5} & 0 \\ 0 & 0 & -\frac{1}{5} \end{pmatrix}, \begin{pmatrix} 0 & 0 & 0 \\ -\frac{\sqrt{3}}{5} & 0 & 0 \\ 0 & \frac{\sqrt{3}}{5} & 0 \end{pmatrix}, \begin{pmatrix} 0 & 0 & 0 \\ 0 & 0 & 0 \\ -\frac{\sqrt{6}}{5} & 0 & 0 \end{pmatrix}\right\}$$

Construct the spin and orbit operators (spin, orbit).

$$\textbf{spin} = \frac{1}{2}\left\{\begin{pmatrix} 0 & 1 \\ 1 & 0 \end{pmatrix}, \begin{pmatrix} 0 & \mathbb{i} \\ -\mathbb{i} & 0 \end{pmatrix}, \begin{pmatrix} -1 & 0 \\ 0 & 1 \end{pmatrix}\right\};$$

$$\textbf{orbit} =$$

$$\frac{1}{\sqrt{2}}\left\{\begin{pmatrix} 0 & \frac{1}{\sqrt{2}} & 0 \\ \frac{1}{\sqrt{2}} & 0 & \frac{1}{\sqrt{2}} \\ 0 & \frac{1}{\sqrt{2}} & 0 \end{pmatrix}, \begin{pmatrix} 0 & \frac{\mathbb{i}}{\sqrt{2}} & 0 \\ -\frac{\mathbb{i}}{\sqrt{2}} & 0 & \frac{\mathbb{i}}{\sqrt{2}} \\ 0 & -\frac{\mathbb{i}}{\sqrt{2}} & 0 \end{pmatrix}, \begin{pmatrix} -1 & 0 & 0 \\ 0 & 0 & 0 \\ 0 & 0 & 1 \end{pmatrix}\right\};$$

Table[MatrixForm[**spin**[[i]]], {i, 1, 3}];
Table[MatrixForm[i[**spinI**[i]]], {i, 1, 3}];
Table[MatrixForm[**orbit**[[i]]], {i, 1, 3}];
Table[MatrixForm[i[**orbitI**[i]]], {i, 1, 3}];

Transition matrix elements for He-4

For 2S states (1 singlet and 3 triplets):  write Hamiltonian and diagonalize

Create operators, s1 and s2.

Electron 1 spin operator (s1), electron 2 spin operator (s2)

**s1** =  Table[**spin**[[$i$]] $\otimes$ IdentityMatrix[2], {$i$, 1, 3}];

**s2** =  Table[IdentityMatrix[2] $\otimes$ **spin**[[$i$]], {$i$, 1, 3}];

Table[MatrixForm[**s1**[[$i$]]], {$i$, 1, 3}]
Table[MatrixForm[**s2**[[$i$]]], {$i$, 1, 3}]

$$\left\{ \begin{pmatrix} 0 & 0 & \frac{1}{2} & 0 \\ 0 & 0 & 0 & \frac{1}{2} \\ \frac{1}{2} & 0 & 0 & 0 \\ 0 & \frac{1}{2} & 0 & 0 \end{pmatrix}, \begin{pmatrix} 0 & 0 & \frac{i}{2} & 0 \\ 0 & 0 & 0 & \frac{i}{2} \\ -\frac{i}{2} & 0 & 0 & 0 \\ 0 & -\frac{i}{2} & 0 & 0 \end{pmatrix}, \begin{pmatrix} -\frac{1}{2} & 0 & 0 & 0 \\ 0 & -\frac{1}{2} & 0 & 0 \\ 0 & 0 & \frac{1}{2} & 0 \\ 0 & 0 & 0 & \frac{1}{2} \end{pmatrix} \right\}$$

$$\left\{ \begin{pmatrix} 0 & \frac{1}{2} & 0 & 0 \\ \frac{1}{2} & 0 & 0 & 0 \\ 0 & 0 & 0 & \frac{1}{2} \\ 0 & 0 & \frac{1}{2} & 0 \end{pmatrix}, \begin{pmatrix} 0 & \frac{i}{2} & 0 & 0 \\ -\frac{i}{2} & 0 & 0 & 0 \\ 0 & 0 & 0 & \frac{i}{2} \\ 0 & 0 & -\frac{i}{2} & 0 \end{pmatrix}, \begin{pmatrix} -\frac{1}{2} & 0 & 0 & 0 \\ 0 & \frac{1}{2} & 0 & 0 \\ 0 & 0 & -\frac{1}{2} & 0 \\ 0 & 0 & 0 & \frac{1}{2} \end{pmatrix} \right\}$$

Introduce interactions, change to S basis

$\Delta s$ = empirical singlet-triplet splitting parameter, gj = electron g-factor, m = Bohr magneton, b = magnetic field value
s1s2 = spin-spin interaction operator, S = total spin operator, K = singlet-triplet mixing operator

nv2s = SetPrecision[{Δs → −192 502 669, gj → 2.002237348, m → 1.39962418, b → **bvalue**}, 17];

$\mathbf{sls2} = \Delta s \sum_{i=1}^{3} \mathbf{s1}[[i]] \cdot \mathbf{s2}[[i]];$

$S = \mathbf{s1} + \mathbf{s2};$

$K = \mathbf{s1} - \mathbf{s2};$

$H = \mathbf{sls2} - \dfrac{1}{4} \Delta s\, \text{IdentityMatrix}[\text{Length}[\mathbf{sls2}]];$

{**Svals**, **Svecs**} = EigenSys[(**sls2** + b S[[3]]) /. {b → 1, Δs → −10}];

$S = \text{Table}\big[\mathbf{Svecs} \cdot S[[i]] \cdot \mathbf{Svecs}^T, \{i, 1, 3\}\big];$

$K = \text{Table}\big[\mathbf{Svecs} \cdot K[[i]] \cdot \mathbf{Svecs}^T, \{i, 1, 3\}\big];$

$H = \mathbf{Svecs} \cdot H \cdot \mathbf{Svecs}^T + b\, \text{gj}\, m\, S[[3]];$

Table[MatrixForm[S[[i]]], {i, 1, 3}]
Table[MatrixForm[K[[i]]], {i, 1, 3}]
MatrixForm[H]

**E2s** = Sort[Eigenvalues[H /. **nv2s**]];

**nv2sB** = Table[SetPrecision[ReplacePart[**nv2s**, **brange**[[i]], Length[**nv2s**]], Precision[**nv2s**]], {i, Length[**brange**]}];
**E2sB** = Table[Sort[Eigenvalues[H /. **nv2sB**[[i]]]], {i, Length[**nv2sB**]}];
TableForm[**E2sB**];

**H2s** = H;

$\left\{ \begin{pmatrix} 0 & \frac{1}{\sqrt{2}} & 0 & 0 \\ \frac{1}{\sqrt{2}} & 0 & \frac{1}{\sqrt{2}} & 0 \\ 0 & \frac{1}{\sqrt{2}} & 0 & 0 \\ 0 & 0 & 0 & 0 \end{pmatrix}, \begin{pmatrix} 0 & \frac{i}{\sqrt{2}} & 0 & 0 \\ -\frac{i}{\sqrt{2}} & 0 & \frac{i}{\sqrt{2}} & 0 \\ 0 & -\frac{i}{\sqrt{2}} & 0 & 0 \\ 0 & 0 & 0 & 0 \end{pmatrix}, \begin{pmatrix} -1 & 0 & 0 & 0 \\ 0 & 0 & 0 & 0 \\ 0 & 0 & 1 & 0 \\ 0 & 0 & 0 & 0 \end{pmatrix} \right\}$

$\left\{ \begin{pmatrix} 0 & 0 & 0 & \frac{1}{\sqrt{2}} \\ 0 & 0 & 0 & 0 \\ 0 & 0 & 0 & -\frac{1}{\sqrt{2}} \\ \frac{1}{\sqrt{2}} & 0 & -\frac{1}{\sqrt{2}} & 0 \end{pmatrix}, \begin{pmatrix} 0 & 0 & 0 & \frac{i}{\sqrt{2}} \\ 0 & 0 & 0 & 0 \\ 0 & 0 & 0 & \frac{i}{\sqrt{2}} \\ -\frac{i}{\sqrt{2}} & 0 & -\frac{i}{\sqrt{2}} & 0 \end{pmatrix}, \begin{pmatrix} 0 & 0 & 0 & 0 \\ 0 & 0 & 0 & 1 \\ 0 & 0 & 0 & 0 \\ 0 & 1 & 0 & 0 \end{pmatrix} \right\}$

$\begin{pmatrix} -b\, \text{gj}\, m & 0 & 0 & 0 \\ 0 & 0 & 0 & 0 \\ 0 & 0 & b\, \text{gj}\, m & 0 \\ 0 & 0 & 0 & -\Delta s \end{pmatrix}$

For 2P states (3 singlets and 9 triplets):  write Hamiltonian and diagonalize

Create operators, s1, s2, and L.

Electron 1 spin operator (s1), electron 2 spin operator (s2) and outer electron angular momentum operator (L)

**s1** =  Table[IdentityMatrix[3] $\otimes$ (**spin**[[$i$]] $\otimes$ IdentityMatrix[2]), {$i$, 1, 3}];
**s2** =  Table[IdentityMatrix[3] $\otimes$ (IdentityMatrix[2] $\otimes$ **spin**[[$i$]]), {$i$, 1, 3}];
$L$ =  Table[**orbit**[[$i$]] $\otimes$ (IdentityMatrix[2] $\otimes$ IdentityMatrix[2]), {$i$, 1, 3}];

Table[MatrixForm[**s1**[[$i$]]], {$i$, 1, 3}];
Table[MatrixForm[**s2**[[$i$]]], {$i$, 1, 3}];
Table[MatrixForm[$L$[[$i$]]], {$i$, 1, 3}];

Introduce spin-spin interactions, change to S basis

Spin-spin interaction operator (s1s2), total spin operator (S), singlet-triplet mixing operator (K)

**s1s2** = $\sum_{i=1}^{3}$ **s1**[[$i$]] $\cdot$ **s2**[[$i$]];

$S$ =  **s1** + **s2**;
$K$ =  **s1** − **s2**;

$H$ =  $\triangle$p **s1s2**;
{**Svalp**, **Svecp**} =  EigenSys[($H$ + b1 $L$[[3]] + b2 $S$[[3]]) /. {b1 $\to$ 1, b2 $\to$ 10, $\triangle$p $\to$ −100}];

**Stl** =  Table[**Svecp** $\cdot$ $S$[[$i$]] $\cdot$ **Svecp**$^T$, {$i$, 1, 3}];
**Ktl** =  Table[**Svecp** $\cdot$ $K$[[$i$]] $\cdot$ **Svecp**$^T$, {$i$, 1, 3}];
**Ltl** =  Table[**Svecp** $\cdot$ $L$[[$i$]] $\cdot$ **Svecp**$^T$, {$i$, 1, 3}];
**Htl** =  **Svecp** $\cdot$ $H$ $\cdot$ **Svecp**$^T$;

Table[MatrixForm[**Stl**[[$i$]]], {$i$, 1, 3}];
Table[MatrixForm[**Ktl**[[$i$]]], {$i$, 1, 3}];
Table[MatrixForm[**Ltl**[[$i$]]], {$i$, 1, 3}];
MatrixForm[**Htl**];

Change to J = L + S basis

Em = (J=1) mixing, Δp = empirical s-t splitting parameter, E0 = ~ He4 0-2, E1'= He4 2-1
w/o s-t mixing, gs = electron g-factor, gl = orbital g-factor, m = Bohr magneton,
b = magnetic field value
LS = spin-orbit interaction operator, LK = singlet-triplet-orbit interaction operator,
J = total angular momentum operator

**nvalues** = SetPrecision[{Em → −17 085.298, Δp → −61 431 000, E0 → 31 908.135, E1′ → 2291.173 + 4.75197,
$\qquad$ gs → 2.002238867, gl → 0.999873626, $m$ → 1.39962418, $b$ → **bvalue**}, 45];

**LS** = ls $\sum_{i=1}^{3}$ **Stl**[[i]] · **Ltl**[[i]];

**LK** = $\sum_{i=1}^{3}$ **Ktl**[[i]] · **Ltl**[[i]];

$J$ = **Ltl** + **Stl**;

{**Svalj**, **Svecj**} = EigenSys[(**Htl** + **LS** + $b$ $J$[[3]]) /. {Δp → −100, ls → −10, $b$ → 1}];
**Svecjp** = **Svecj** · **Svecp**;

**Jt** = Table[**Svecj** · $J$[[i]] · **Svecj**$^T$, {i, 1, 3}];

**LKt2** = **Svecj** · **LK** · **Svecj**$^T$;

**Ht2** = $\dfrac{\text{Em }\mathbf{LKt2}}{\sqrt{2}}$ + DiagonalMatrix[{0, 0, 0, 0, 0, E1′, E1′, E1′, E0, 0, 0, 0}] + **Svecj** · **Htl** · **Svecj**$^T$

$\qquad - \dfrac{1}{4}$ Δp IdentityMatrix[Length[**Svecj**]];

**Ht3** = **Svecj**$^T$ · **Ht2** · **Svecj**;

Table[MatrixForm[**Jt**[[i]]], {i, 1, 3}];
MatrixForm[**LKt2**];
MatrixForm[**Ht2**];
MatrixForm[**Ht3**];

**Hb** = $m$ $b$ (gl **Ltl**[[3]] + gs **Stl**[[3]]);
{**SvalH**, **SvecH**} = EigenSys (**Hb** + **Ht3** /. **nvalues**);
**E2p** = **SvalH**;
**SvecHp** = **SvecH** · **Svecp**;

Resort eigenvalues and eigenvectors, necessary when levels cross at high B-field.

**PBOrder** = Extract[Flatten[Table[Sort[Transpose[{1, 2, 3, 4, 5, 6, 7, 8, 9, 0, 0, 0}Transpose[

$$Round\left[\begin{pmatrix} 1 & 0 & 0 & 0 & 0 & 0 & 0 & 0 & 0 & 0 & 0 & 0 \\ 0 & 1 & 0 & 1 & 0 & 0 & 0 & 0 & 0 & 0 & 0 & 0 \\ 0 & 0 & 1 & 0 & 1 & 0 & 1 & 0 & 0 & 0 & 0 & 0 \\ 0 & 0 & 0 & 0 & 0 & 1 & 0 & 1 & 0 & 0 & 0 & 0 \\ 0 & 0 & 0 & 0 & 0 & 0 & 0 & 1 & 0 & 0 & 0 \end{pmatrix}.Transpose[\mathbf{SvecH}]\textasciicircum 2\right]\right][[i]], Greater\right], \{i, 1, 5\}\right],$$

    {{1}, {13}, {14}, {25}, {26}, {27}, {37}, {38}, {49}}];

**PLBOrder** = {1, 6, 2, 9, 7, 3, 8, 4, 5};
**Pmj** = {, , , , , , , , };
Do[**Pmj**[[**PLBOrder**[[i]]]] = **PBOrder**[[i]], {i, 1, 9}];

MatrixForm$\left[Chop\left[Simplify\left[\mathbf{SvecH}\cdot(\mathbf{Hb}+\mathbf{Ht3}\ /.\ \mathbf{nvalues})\cdot\mathbf{SvecH}^T\right],\ \dfrac{1}{10^{16}}\right]\right]$;

**nv2pB** = Table[SetPrecision[ReplacePart[**nvalues**, **brange**[[i]], Length[**nvalues**]], Precision[**nvalues**]],
      {i, Length[**brange**]}];
**E2pB** = Table[Sort[Eigenvalues[**Hb**+**Ht3**/.**nv2pB**[[i]]]], {i, Length[**nv2pB**]}];
TableForm[**E2pB**];

Transform electric dipole operator

**edmsml** = Table[
      **edm**[[i]]⊗(IdentityMatrix[2]⊗IdentityMatrix[2]), {i, 1, 3}];
Table[MatrixForm[**edmsml**[[i]]], {i, 1, 3}];
**Svec** = BlockMatrix[{{**Svecs**, ConstantArray[0, {Length[**Svecs**], Length[**Svecjp**]}]},
      {ConstantArray[0, {Length[**Svecjp**], Length[**Svecs**]}], **Svecjp**}}];
**edj** = Table[**Svec**·**edmsml**[[i]]·Transpose[**Svec**], {i, 1, 3}];
Table[MatrixForm[**edj**[[i]]], {i, 1, 3}];
**t1** = vec[**edj**];
**edsq** = Table[**t1**[[i, j, k]]**t1**[[i, j, k]]*, {i, 1, 3}, {j, 1, Sq}, {k, 1, Sq}];
Table[MatrixForm[**edsq**[[i]]], {i, 1, 3}];
**Svec** = BlockMatrix[{{**Svecs**, ConstantArray[0, {Length[**Svecs**], Length[**SvecHp**]}]},
      {ConstantArray[0, {Length[**SvecHp**], Length[**Svecs**]}], **SvecHp**}}];
**edH** = Table[**Svec**·**edmsml**[[i]]·Transpose[**Svec**], {i, 1, 3}];
Table[MatrixForm[**edH**[[i]]], {i, 1, 3}];
**t2** = vec[**edH**];
**edHsq** = N$\left[Chop\left[Table[\mathbf{t2}[[i, j, k]]\mathbf{t2}[[i, j, k]]^*, \{i, 1, 3\}, \{j, 1, Length[\mathbf{t2}[[1]]]\}, \{k, 1, Length[\mathbf{t2}[[1]]]\}\right],\right.$
      $\left.\left.\dfrac{1}{10^{10}}\right], 10\right]$;
Table[MatrixForm[**edHsq**[[i]]], {i, 1, 3}];

Determine rates into upper states, decay rates into other states, and signal size, repopulation rates, branching into detectable states, frequency

**E2** =  Join[**E2s**, **E2p**];

Table$\left[\sum_{i=1}^{3}\sum_{j=1}^{16}\textbf{edHsq}[[i, k, j]], \{k, 16\}\right]$;

**t1** =  Table[i, {i, 5, 13}];

**t2** = Table[{**edHsq**[[3, Pmj[[i]] + 4, j]], **edHsq**[[1, Pmj[[i]] + 4, j]],

    If[j == 2, −2 **edHsq**[[1, Pmj[[i]] + 4, 2]] −**edHsq**[[3, Pmj[[i]] + 4, 2]] +1,

      2 **edHsq**[[1, Pmj[[i]] + 4, 2]] +**edHsq**[[3, Pmj[[i]] + 4, 2]]],

    (**edHsq**[[1, Pmj[[i]] + 4, j]] +**edHsq**[[3, Pmj[[i]] + 4, j]])

      If[j == **2**, −2 **edHsq**[[1, Pmj[[i]] + 4, 2]] −**edHsq**[[3, Pmj[[i]] + 4, 2]] +1,

      2 **edHsq**[[1, Pmj[[i]] + 4, 2]] +**edHsq**[[3, Pmj[[i]] + 4, 2]]],

    (**edHsq**[[1, Pmj[[i]] + 4, j]] +**edHsq**[[3, Pmj[[i]] + 4, j]])

      (2 **edHsq**[[1, Pmj[[i]] + 4, j]] +**edHsq**[[3, Pmj[[i]] + 4, j]]),

    If[j == 2, Null, (2 **edHsq**[[1, Pmj[[i]] + 4, j]] +**edHsq**[[3, Pmj[[i]] + 4, j]]) /

      (2 **edHsq**[[1, Pmj[[i]] + 4, 1]] +2 **edHsq**[[3, Pmj[[i]] + 4, 3]] +**edHsq**[[3, Pmj[[i]] + 4, 1]]

        +**edHsq**[[3, Pmj[[i]] + 4, 3]])], **E2**[[**Pmj**[[i]] +4]] −**E2**[[,j]]], {i, 1, 9}, {j, 1, 3}];

**Signal** =  Table[**t2**[[i, j, k]], {i, 9}, {k, {1, 2, 3, 4, 5, 7}}, {j, 3}];

Style[NumberForm[TableForm[

    Table[Prepend[Round[**Signal**[[i]], .001], {"−1", "0", "+1"}], {i, Length[Signal]}], TableAlignments → Center,

    TableHeadings → {{"J2(−2)", "J2(−1)", "J2(0)", "J2(+1)", "J2(+2)", "J1(−1)", "J1(0)", "J1(+1)", "J0(0)"},

      {"2S", " E∥B ", "E⊥B", "det.decay", "signal", "repopulate", "frequency"}}], {5, 3}],

  FontSize → 12]

| | 2S | E∥B | E⊥B | det.decay | signal | repopulate | frequency |
|---|---|---|---|---|---|---|---|
| | -1 | 0.000 | 0.500 | 0.000 | 0.000 | 0.500 | -55.978 |
| | 0 | 0.000 | 0.000 | 1.000 | 0.000 | 0.000 | -168.070 |
| J2(-2) | +1 | 0.000 | 0.000 | 0.000 | 0.000 | 0.000 | -280.170 |
| | -1 | 0.512 | 0.000 | 0.488 | 0.250 | 0.262 | 27.715 |
| | 0 | 0.000 | 0.244 | 0.512 | 0.125 | 0.119 | -84.380 |
| J2(-1) | +1 | 0.000 | 0.000 | 0.488 | 0.000 | 0.000 | -196.480 |
| | -1 | 0.000 | 0.087 | 0.667 | 0.058 | 0.015 | 111.640 |
| | 0 | 0.667 | 0.000 | 0.333 | 0.222 | 0.444 | -0.458 |
| J2(0) | +1 | 0.000 | 0.079 | 0.667 | 0.053 | 0.013 | -112.550 |
| | -1 | 0.000 | 0.000 | 0.512 | 0.000 | 0.000 | 195.790 |
| | 0 | 0.000 | 0.256 | 0.488 | 0.125 | 0.131 | 83.693 |
| J2(+1) | +1 | 0.488 | 0.000 | 0.512 | 0.250 | 0.238 | -28.402 |
| | -1 | 0.000 | 0.000 | 0.000 | 0.000 | 0.000 | 280.170 |
| | 0 | 0.000 | 0.000 | 1.000 | 0.000 | 0.000 | 168.070 |
| J2(+2) | +1 | 0.000 | 0.500 | 0.000 | 0.000 | 0.500 | 55.978 |
| | -1 | 0.488 | 0.000 | 0.512 | 0.250 | 0.238 | 2319.600 |
| | 0 | 0.000 | 0.256 | 0.488 | 0.125 | 0.131 | 2207.500 |
| J1(-1) | +1 | 0.000 | 0.000 | 0.512 | 0.000 | 0.000 | 2095.400 |
| | -1 | 0.000 | 0.247 | 0.000 | 0.000 | 0.122 | 2403.700 |
| | 0 | 0.000 | 0.000 | 1.000 | 0.000 | 0.000 | 2291.600 |
| J1(0) | +1 | 0.000 | 0.253 | 0.000 | 0.000 | 0.128 | 2179.500 |
| | -1 | 0.000 | 0.000 | 0.488 | 0.000 | 0.000 | 2487.600 |
| | 0 | 0.000 | 0.244 | 0.512 | 0.125 | 0.119 | 2375.600 |
| J1(+1) | +1 | 0.512 | 0.000 | 0.488 | 0.250 | 0.262 | 2263.500 |
| | -1 | 0.000 | 0.166 | 0.333 | 0.055 | 0.055 | 32020.000 |
| | 0 | 0.333 | 0.000 | 0.667 | 0.222 | 0.111 | 31908.000 |
| J0(0) | +1 | 0.000 | 0.167 | 0.333 | 0.056 | 0.056 | 31796.000 |

Fit B-Field corrections to polynomials

**bvalues** = Table[SetPrecision[$b$ /. **brange**[[$i$]], Precision[**E2pB**]], {$i$, Length[**brange**]}];
**Bdata** = Flatten[Table[
    {**bvalues**[[$i$]], **E2pB**[[$i$, $j$]] −**E2sB**[[$i$, $k$]] −If[**bvalues**[[1]] == 0, **E2pB**[1, $j$] −**E2sB**[1, 1]]},
    {$j$, Length[**E2pB**[[1]]]}, {$k$, 3}, {$i$, Length[**bvalues**]}], 1];
**Btrans** = {27, 26, 25, 24, 23, 21, 19, 17, 16, 12, 11, 9, 8, 7, 5, 4};
**Bfit** = Table[Fit[**Bdata**[[**Btrans**[[$i$]]]], {$E$, $E^2$, $E^3$, $E^4$}, $E$], {$i$, Length[**Btrans**]}];
Style[NumberForm[TableForm[Table[
    {{"J0m0(+1)    ", "J0m0(0)", "J0m0(−1)", "J1m+1(+1)", "J1m+1(0)", "J1m0(+1)", "J1m0(−1)",
      "J1m−1(0)", "J1m−1(−1)", "J2m+1(+1)", "J2m+1(0)", "J2m0(+1)", "J2m0(0)",
      "J2m0(−1)", "J2m−1(0)", "J2m−1(−1)"}[[$i$]], **Bfit**[[$i$]]}, {$i$, Length[**Bfit**]}]], 4], FontSize → 10]
**Blfit** = Table[LinearFit[(**Bdata**[[**Btrans**[[$i$]]]]), {1, 2, 3, 4}, $b$), {$i$, Length[**Btrans**]}];

| | |
|---|---|
| J0m0(+1) | $-2.802\,B + 0.00004430\,B^2 - \left(2.\times 10^{-16}\right) B^3 - \left(3.552 \times 10^{-14}\right) B^4$ |
| J0m0(0) | $\left(-7.243 \times 10^{-14}\right) B + 0.00004430\,B^2 - \left(1.674 \times 10^{-16}\right) B^3 - \left(3.552 \times 10^{-14}\right) B^4$ |
| J0m0(-1) | $2.802\,B + 0.00004430\,B^2 - \left(2.\times 10^{-16}\right) B^3 - \left(3.552 \times 10^{-14}\right) B^4$ |
| J1m+1(+1) | $-0.7015\,B + 0.0002148\,B^2 - \left(1.616 \times 10^{-11}\right) B^3 - \left(1.999 \times 10^{-11}\right) B^4$ |
| J1m+1(0) | $2.101\,B + 0.0002148\,B^2 - \left(1.616 \times 10^{-11}\right) B^3 - \left(1.999 \times 10^{-11}\right) B^4$ |
| J1m0(+1) | $-2.802\,B + 0.0002420\,B^2 - \left(3.023 \times 10^{-11}\right) B^3 - \left(3.034 \times 10^{-11}\right) B^4$ |
| J1m0(-1) | $2.802\,B + 0.0002420\,B^2 - \left(3.023 \times 10^{-11}\right) B^3 - \left(3.034 \times 10^{-11}\right) B^4$ |
| J1m-1(0) | $-2.101\,B + 0.0002148\,B^2 - \left(1.617 \times 10^{-11}\right) B^3 - \left(1.999 \times 10^{-11}\right) B^4$ |
| J1m-1(-1) | $0.7015\,B + 0.0002148\,B^2 - \left(1.617 \times 10^{-11}\right) B^3 - \left(1.999 \times 10^{-11}\right) B^4$ |
| J2m+1(+1) | $-0.7015\,B - 0.0002148\,B^2 + \left(1.616 \times 10^{-11}\right) B^3 + \left(1.999 \times 10^{-11}\right) B^4$ |
| J2m+1(0) | $2.101\,B - 0.0002148\,B^2 + \left(1.616 \times 10^{-11}\right) B^3 + \left(1.999 \times 10^{-11}\right) B^4$ |
| J2m0(+1) | $-2.802\,B - 0.0002864\,B^2 + \left(3.023 \times 10^{-11}\right) B^3 + \left(3.038 \times 10^{-11}\right) B^4$ |
| J2m0(0) | $\left(1.307 \times 10^{-8}\right) B - 0.0002864\,B^2 + \left(3.023 \times 10^{-11}\right) B^3 + \left(3.038 \times 10^{-11}\right) B^4$ |
| J2m0(-1) | $2.802\,B - 0.0002864\,B^2 + \left(3.023 \times 10^{-11}\right) B^3 + \left(3.038 \times 10^{-11}\right) B^4$ |
| J2m-1(0) | $-2.101\,B - 0.0002148\,B^2 + \left(1.617 \times 10^{-11}\right) B^3 + \left(1.999 \times 10^{-11}\right) B^4$ |
| J2m-1(-1) | $0.7015\,B - 0.0002148\,B^2 + \left(1.617 \times 10^{-11}\right) B^3 + \left(1.999 \times 10^{-11}\right) B^4$ |

Plot eigenvalues vs. B-field

**bvalues** = Table[1 + i 10, {i, 0, 1000}];
**nv2sBs** = Table[SetPrecision[{Δs → −192 502 669, gj → 2.002237348`, m → 1.39962418`, b → **bvalues**[[i]]},
    25], {i, Length[**bvalues**]}];
**nv2pBs**
    = Table[SetPrecision[{Em → −17085.298`, Δp → −61 431 000, E0 → 31908.135`,
        E1′ → 2291.173` + 4.75197`, gs → 2.002238867`, gl → 0.999873626`, m → 1.39962418`,
        b → **bvalues**[[i]]}, 25], {i, Length[**bvalues**]}];

**Pvs** = Table[EigenSys[(**Hb** + **Ht3**) /. **nv2pBs**[[i]]], {i, Length[**bvalues**]}];
**PBOrder** = Table[Extract[Flatten[Table[Sort[
        Transpose[{1, 2, 3, 4, 5, 6, 7, 8, 9, 0, 0, 0}Transpose[
$$
\text{Round}\left[\begin{pmatrix} 1 & 0 & 0 & 0 & 0 & 0 & 0 & 0 & 0 & 0 & 0 & 0 \\ 0 & 1 & 0 & 1 & 0 & 0 & 0 & 0 & 0 & 0 & 0 & 0 \\ 0 & 0 & 1 & 0 & 1 & 0 & 1 & 0 & 0 & 0 & 0 & 0 \\ 0 & 0 & 0 & 0 & 0 & 1 & 0 & 1 & 0 & 0 & 0 & 0 \\ 0 & 0 & 0 & 0 & 0 & 0 & 0 & 1 & 0 & 1 & 0 & 0 \end{pmatrix}.\text{Transpose}[\textbf{Pvs}[[b, 2]]]^2\right]\right][[i]], \text{Greater}], \{i, 1, 5\}\Big],
$$
        {{1}, {13}, {14}, {25}, {26}, {27}, {37}, {38}, {49}}], {b, 1, Length[**bvalues**]}];
**PLBOrder** = {1, 6, 2, 9, 7, 3, 8, 4, 5};
**Pmap** = Table[{Null, Null, Null, Null, Null, Null, Null, Null, Null}, {b, 1, Length[**bvalues**]}];
Do[**Pmap**[[b, **PLBOrder**[[i]]]] = **PBOrder**[[b, i]], {b, 1, Length[**bvalues**]}, {i, 1, 9}];

**Bdata2p** = Table[**Pvs**[[b, 1, **Pmap**[[b, i]]]], {i, 1, 9}, {b, 1, Length[**bvalues**]}];
**Bdata2s** = Transpose[Table[Sort[Eigenvalues[**H2s** /. **nv2sBs**[[i]]]], {i, Length[**bvalues**]}]];
**BdataTrans** = Flatten[Table[**Bdata2p**[[i]] − **Bdata2s**[[j]], {i, Length[**Bdata2p**]}, {j, Length[**Bdata2s**]}], 1];

**SLevels** = {−1, 0, 1};
**PLevels** = {−2, −1, 0, 1, 2, −1, 0, 1, 0};

Plots
    = Table[
      ListPlot[If[Abs[**PLevels**[[i]] − **SLevels**[[j]]] ≤ 1,
          Table[{**bvalues**[[b]], **Bdata2p**[[i, b]] − **Bdata2s**[[j, b]]}, {b, Length[**bvalues**]}], {0, 0}],
          Joined → True, DisplayFunction → Identity,
          PlotStyle → {RGBColor[If[Abs[**PLevels**[[i]] − **SLevels**[[j]]] == 0, 0, 1], 0,
              If[Abs[**PLevels**[[i]] − **SLevels**[[j]]] == 0, 1, 0]]}], {i, 1, 9}, {j, 1, 3}];
Show[Table[**Plots**[[i]], {i, Length[**Plots**]}], PlotRange → {{0, 10000}, {−10 000, 50 000}},
  DisplayFunction → $DisplayFunction]
**Plots** = Table[ListPlot[Table[{**bvalues**[[b]], **Bdata2p**[[i, b]]}, {b, Length[**bvalues**]}], Joined → True,
      DisplayFunction → Identity], {i, Length[**Bdata2p**]}];
Show[Table[**Plots**[[i]], {i, Length[**Plots**]}], PlotRange → {{0, 10 000}, {−1000, 35 000}},
  DisplayFunction → $DisplayFunction]

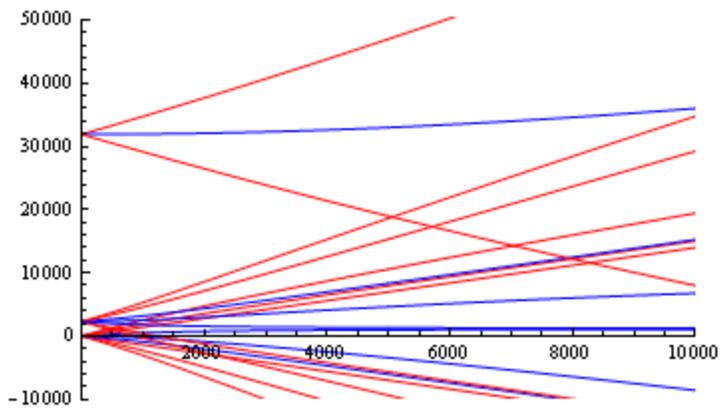

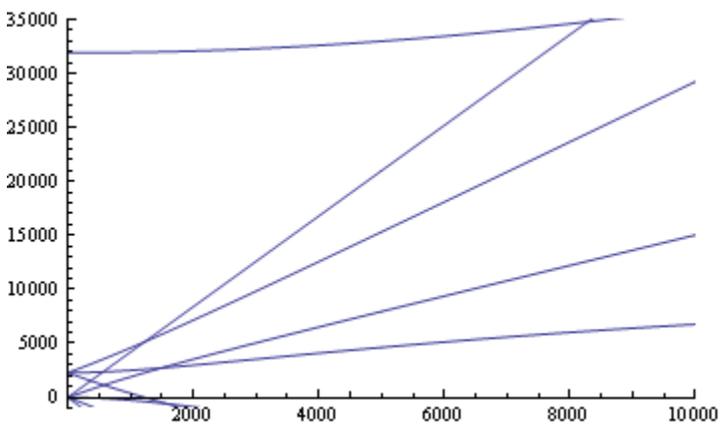

View He-4 Transitions, bold transitions are from m = 0 initial states, light transitions are from m = ±1 initial states.

**fr** = {0, 2291.173, 31 908.135};

**tl** = Table[Plot[Evaluate[lorz(**Signal**[[i, 4, j]], **Signal**[[i, 6, j]], f + **fr**[[fs]])], {f, −150, 150},

PlotRange → {0, .25}, AspectRatio → $\frac{1}{5}$, AxesOrigin → {−150, 0}, DisplayFunction → Identity,

PlotStyle → {RGBColor[If[**Signal**[[i, 1, j]] > **Signal**[[i, 2, j]], 0, 1], 0,

If[**Signal**[[i, 1, j]] > **Signal**[[i, 2, j]], 1, 0]], AbsoluteThickness[If[j == 2, 2, 1]]}],

{fs, 1, 3}, {i, 1, 9}, {j, 1, 3}];

Show[**tl**[[1]], DisplayFunction → $DisplayFunction, PlotRange → {0, .25}]

Show[**tl**[[2]], DisplayFunction → $DisplayFunction, PlotRange → {0, .25}]

Show[**tl**[[3]], DisplayFunction → $DisplayFunction, PlotRange → {0, .25}]

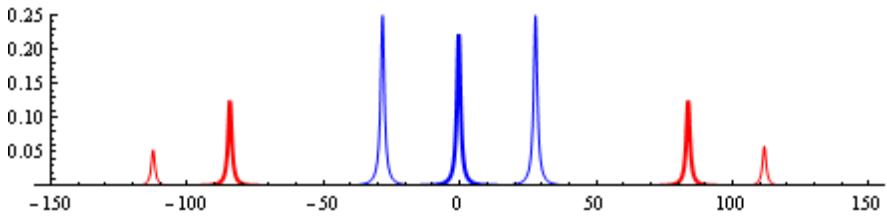

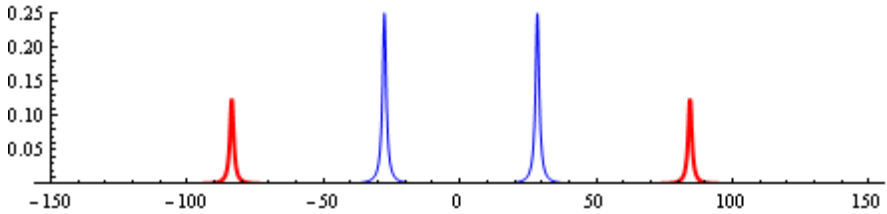

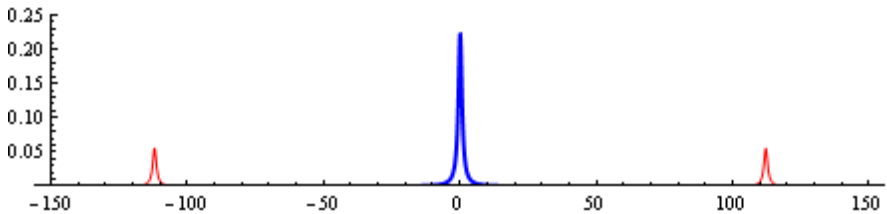

REFERENCES

1. D. Hanneke, S. Fogwell, and G. Gabrielse, Phys. Rev. Lett. 100, 120801 (2008).

2. H. A. Bethe and E. E. Salpeter, *Quantum Mechanics of One- and Two-Electron Atoms*, Springer-Verlag, 1957.

3. C. Schwartz, Phys. Rev. 134, A1181 (1964).

4. F. M. J. Pichanick, R. D. Swift, C. E. Johnson, and V. W. Hughes, Phys. Rev. 169, 55 (1968).

5. M. H. Douglas and N. M. Kroll, Ann. Phys. (NY) 82, 89 (1974).

6. M. L. Lewis and P. H. Serafino, Phys. Rev. A 18, 867 (1978).

7. T. Zhang, Z. C. Yan, and G. W. F. Drake, Phys. Rev. Lett. 77, 1715 (1996).

8. K. Pachucki and J. Sapirstein, Phys. Rev. Lett. 97, 013002 (2006).

9. K. Pachucki and V. A. Yerokhin, Phys. Rev. Lett. 104, 070403 (2010).

10. M. C. George, L. D. Lombardi, and E. A. Hessels, Phys. Rev. Lett. 87, 173002 (2001).

11. J. S. Borbely, M. C. George, L. D. Lombardi, M. Weel, D. W. Fitzakerley, and E. A. Hessels, Phys. Rev. A 79, 060503(R) (2009).

12. J. Castillega, D. Livingston, A. Sanders, and D. Shiner, Phys. Rev. Lett. 84, 4321 (2000).

13. D. Shiner, R. Dixson, IEEE Trans. Instrum. Meas. 44, 518 (1995).

14. T. Zelevinsky, D. Farkas, and G. Gabrielse, Phys. Rev. Lett. 95, 203001 (2005).

15. G. Giusfredi, P. Cancio Pastor, P. De Natale, D. Mazzotti, C. de Mauro, L. Fallani, G. Hagel, V. Krachmalnicoff, and M. Inguscio, Can. J. Phys. 83, 301 (2005).

16. D. J. Griffiths, *Introduction to Quantum Mechanics*, Prentice Hall, Upper Saddle River, NJ, 1995.

17. R. Shankar, Principles of Quantum Mechanics, Plenum Press, New York , NY 1994.

18. M. L. Lewis and V. W. Hughes, Phys. Rev. A 8, 2845 (1973).

19. H. K. Holt and R. Krotkov, Phys. Rev. 144, 82 (1966).

20. F. Rosebury, *Handbook of Electron Tube and Vacuum Techniques*, Addison Wesley, Reading, MA, 1965.

21. R. S. Van Dyck, Jr., C. E. Johnson, H. A. Shugart, Phys. Rev. A 4, 1327 (1971).

22. C. E. Johnson, Phys. Rev. A 7, 872 (1973).

23. D. J. Johnson, H. C. Ives, M. E. Savage, and W. A. Stygar, "High-Voltage Hold-Off of Large Surface Dielectric Surface Layers," Pulsed Power Conference, 2003. Digest of Technical Papers. PPC-2003. 14th IEEE International.

24. S. D. Rosner and F. M. Pipkin, Phys. Rev. A 1, 571 (1970).

25. K. Pachucki and V. A. Yerokhin, Phys. Rev. A 79, 062516 (2009).

26. K. Pachucki and V. A. Yerokhin, Phys. Rev. A 81, 039903(E) (2010).

27. K. Pachucki and J. Sapirstein, J. Phys. B: At. Mol. Opt. Phys. 35, 1783 (2002).

28. M. Cadoret, E. de Mirandes, P. Clade, S. Guellati-Khelifa, C. Schwob, F. Nez, L. Julien, and F. Biraben, Phys. Rev. Lett. 101, 230801 (2008).

29. V. Gerginov, K. Calkins, C. E. Tanner, J. J. McFerran, S. Diddams, A. Bartels, and L. Hollberg, Phys. Rev. A 73, 032504 (2006).